\documentclass[12pt]{article}
\usepackage{amssymb,amsmath,epsfig}
\allowdisplaybreaks
\begin{document}
\title{\bf Wormhole Geometry and Noether Symmetry in $f(R)$ Gravity}

\author{M. Sharif \thanks{msharif.math@pu.edu.pk}~\thanks{Fellow, Pakistan Academy of Sciences,
3 Constitution Avenue, G-5/2, Islamabad.} and Iqra Nawazish
\thanks{iqranawazish07@gmail.com}\\
Department of Mathematics, University of the Punjab,\\
Quaid-e-Azam Campus, Lahore-54590, Pakistan.}

\date{}

\maketitle
\begin{abstract}
This paper investigates the geometry of static traversable wormhole
through Noether symmetry approach in $f(R)$ gravity. We take perfect
fluid distribution and formulate symmetry generators with associated
conserved quantities corresponding to general form, power-law and
exponential $f(R)$ models. In each case, we evaluate wormhole
solutions using constant and variable red-shift functions. We
analyze the behavior of shape function, viability of constructed
$f(R)$ model and stability of wormhole solutions graphically. The
physical existence of wormhole solutions can be examined through
null/weak energy conditions of perfect fluid and null energy
condition of the effective energy-momentum tensor. The graphical
interpretation of constructed wormhole solutions ensures the
existence of physically viable and traversable wormholes for all
models. It is concluded that the constructed wormholes are found to
be stable in most of the cases.
\end{abstract}

\section{Introduction}

On the landscape of theoretical and observational modern cosmology,
the most revolutionizing fact is believed to be the current cosmic
accelerated expansion. Recent experiments indicate that this
expansion must be due to some enigmatic force with astonishing
anti-gravitational effects, known as dark energy. There are many
proposals to explain its ambiguous nature. The $f(R)$ gravity is one
of such proposals established by replacing geometric part of the
Einstein-Hilbert action with this generic function depending on the
Ricci scalar $R$. The fourth order non-linear field equations of
this gravity keep triggering researchers to evaluate exact solution.

The study of exact solutions under assorted scenarios is extensively
used to explore different cosmic aspects that unveil sophisticated
picture of cosmic evolution. Sharif and Shmair\cite{I1} constructed
vacuum as well as non-vacuum exact solutions of Bianchi I and V
universe models in $f(R)$ gravity and also investigated physical
behavior of these solutions. Guti$\acute{e}$rrez-Pi$\tilde{n}$eres
and L$\acute{o}$pez-Monsalvo \cite{I2} evaluated exact vacuum
solution for static axially symmetric spacetime in the same gravity
and found that solution corresponds to naked singularity. Sharif and
Zubair \cite{11} considered interaction of matter with geometry to
formulate some exact solutions of Bianchi I model. Gao and Shen
\cite{I3} found a new method to formulate exact solutions of static
spherically symmetric metric. They also analyzed some general
properties of solutions like event horizon, singularity and deficit
angle in Jordan and Einstein frames.

Noether symmetry approach is considered to be the most appreciable
technique which explores not only exact solutions but also evaluates
conserved quantities relative to symmetry generators associated with
dynamical system. Capozziello et al. \cite{20} formulated exact
solution of static spherically symmetric metric for $f(R)$ power-law
model. The same authors \cite{I4} generalized this work for
non-static spherically symmetric spacetime and also discussed
possible solutions for axially symmetric model. Vakili \cite{22}
studied the scalar field scenario of flat FRW model through this
approach and discussed current cosmic phase via effective equation
of state parameter corresponding to quintessence phase. Momeni et
al. \cite{a3} investigated the existence of Noether symmetry for
isotropic universe model in mimetic $f(R)$ as well as $f(R,T)$
gravity theories ($T$ denotes trace of energy-momentum tensor).
Sharif and his collaborators \cite{I5} investigated cosmic evolution
as well as current cosmic expansion through Noether symmetry
approach.

Our universe always bring eye opening questions for cosmologists
regrading its surprising and mysterious nature. The existence of
hypothetical geometries is considered as the most debatable issue
which leads to wormhole geometry. A wormhole (WH) structure is
defined through a hypothetical bridge or tunnel which allows a
smooth connection among different regions only if there exists
exotic matter (matter with negative energy density). The existence
of a physically viable WH is questioned due to the presence of
enough amount of exotic matter. Consequently, there is only one way
to have a realistic WH model, i.e., the presence of exotic matter
must be minimized. Besides the existence of such astrophysical
configurations, the most crucial problem is stability analysis which
defines their behavior against perturbations as well as enhances
physical characterization. A singularity-free configuration
identifies a stable state which successfully prevents the WH to
collapse while a WH can also exist for quite a long time even if it
is unstable due to very slow decay. The evolution of unstable system
can lead to many phenomena of interest from structure formation to
supernova explosions. To explore WH existence, different approaches
have been proposed such as modified theories of gravity, non-minimal
curvature-matter coupling, scalar field models etc \cite{I7}.

The study of WH solutions has been of great interest in modified
theories of gravity. Lobo and Oliveira \cite{I6} considered constant
shape function and different fluids to explore WH solution in $f(R)$
gravity. Jamil et al. \cite{I10} formulated viable WH solutions for
$f(R)$ power-law model and also considered particular shape function
in the background of non-commutative geometry. Bahamonde et al.
\cite{I8} constructed cosmological WH threaded by perfect fluid
approaching to FRW universe in the same gravity. Mazharimousavi and
Halilsoy \cite{I9} found a near-throat WH solution of $f(R)$ model
admitting polynomial expansion and also satisfying necessary WH
conditions for both vacuum as well as non-vacuum cases. Sharif and
Fatima \cite{I11} discussed static spherically symmetric WH in
galactic halo region as well as investigated non-static conformal WH
in $f(\mathcal{G})$ gravity, ($\mathcal{G}$ represents Gauss-Bonnet
term). Noether symmetry approach elegantly explores the WH geometry
by formulating exact solutions. Bahamonde et al. \cite{I12} obtained
exact solutions of red-shift as well as shape functions through this
approach and analyzed their geometric behavior graphically in
scalar-tensor theory incorporating non-minimal coupling with torsion
scalar.

In this paper, we study WH geometry threaded by perfect fluid via
Noether symmetry approach in $f(R)$ gravity. The format of the paper
is as follows. Section \textbf{2} explores basic review of $f(R)$
gravity. In section \textbf{3}, we construct point-like Lagrangian
which is used in section \textbf{4} to evaluate WH solutions for
both constant as well as variable red-shift functions. Section
\textbf{5} investigates stability of the constructed WH solutions.
In the last section, we present final remarks.

\section{Basics of $f(R)$ Gravity}

We consider a minimally coupled action of $f(R)$ gravity given by
\begin{equation}\label{1}
\mathcal{I}=\int
d^4x\sqrt{-g}[\frac{f(R)}{2\kappa^2}+\mathcal{L}_m],
\end{equation}
where $g$ identifies determinant of the metric tensor $g_{\mu\nu}$,
$f(R)$ describes a coupling-free function while $\mathcal{L}_m$
denotes Lagrangian density of matter. The metric variation of action
(\ref{1}) leads to
\begin{equation}\label{2}
f_RR_{\mu\nu}-\frac{1}{2}fg_{\mu\nu}-\nabla_\mu\nabla_\nu
f_R+g_{\mu\nu}\Box f_R=\kappa^2T^{(m)}_{\mu\nu},\quad
T^{(m)}_{\mu\nu}=g_{\mu\nu}\mathcal{L}_m-2\frac{\partial\mathcal{L}_m}{\partial
g^{\mu\nu}}.
\end{equation}
Here, $f_R$ shows the derivative of generic function $f$ with
respect to $R$, $\nabla_\mu$ represents covariant derivative,
$\Box=\nabla_{\mu}\nabla^{\mu}$ and $T^{(m)}_{\mu\nu}$ denotes
energy-momentum tensor. The equivalent form of Eq.(\ref{2}) is
\begin{equation}\label{3}
G_{\mu\nu}=\frac{1}{f_R}(T^{(m)}_{\mu\nu}+T^{(c)}_{\mu\nu})=T^{eff}_{\mu\nu},
\end{equation}
where $G_{\mu\nu},~T^{(c)}_{\mu\nu}$ and $T^{eff}_{\mu\nu}$ identify
Einstein, curvature and effective energy-momentum tensors,
respectively. The curvature terms relative to generic function
define $T^{(c)}_{\mu\nu}$ as
\begin{equation}\label{4}
T^{(c)}_{\mu\nu}=\frac{f-Rf_R}{2}g_{\mu\nu}+\nabla_\mu\nabla_\nu
f_R-\Box f_R g_{\mu\nu}.
\end{equation}
The energy-momentum tensor corresponding to perfect fluid is
\begin{equation*}
T^{(m)}_{\mu\nu}=(\rho_m(r)+p_m(r))u_\mu u_\nu+p_m(r)g_{\mu\nu},
\end{equation*}
where $\rho_m$ and $p_m$ characterize energy density and pressure,
respectively whereas $u_\mu$ denotes four velocity of the fluid as
$u_\mu=(-e^{\frac{a(r)}{2}},0,0,0)$.

The static spherically symmetric spacetime is \cite{I13}
\begin{equation}\label{6}
ds^2=-e^{a(r)}dt^2+e^{b(r)}dr^2+M(r)(d\theta^2+\sin^2\theta
d\phi^2),
\end{equation}
where $a,~b$ and $M$ are arbitrary functions depending on radial
coordinate $r$. The geodesic deviation equation determines that
$M(r)=r^2,~\sin r,~\sinh r$ for $\mathcal{K}=0,1,-1$ ($\mathcal{K}$
denotes curvature parameter) under the limiting behavior
$M(r)\rightarrow0$ as $r\rightarrow0$, respectively \cite{I18}. In
case of $M(r)=r^2$, the spherical symmetry defines Morris-Thorne WH
where $a(r)$ is recognized as red-shift function identifying
gravitational red-shift while $e^{b(r)}$ explores the geometry of WH
for $e^b=\left(1-\frac{h(r)}{r}\right)^{-1}$, $h(r)$ is known as
shape function. In order to locate throat of a WH, radial coordinate
must follow non-monotonic behavior such that it decreases from
maximum to minimum value $r_0$ identifying WH throat at $h(r_0)=r_0$
and then it starts increasing from $r_0$ to infinity. To have a WH
solution at throat, the condition $h'(r_0)<1$ is imposed, where
prime denotes derivative with respect to $r$. The flaring-out
condition is the fundamental property of WH which demands
$\frac{h(r)-h(r)'r}{h(r)^2}>0$. For the existence of traversable WH,
the surface should be free from horizons, the red-shift function
must be finite everywhere and $1-h(r)/r>0$. To formulate the field
equations for the action (\ref{1}), we choose $\mathcal{L}_m=p_m(r)$
\cite{I14} and use Eqs.(\ref{2})-(\ref{6}), it follows that
\begin{eqnarray}\nonumber
&&\frac{e^a}{4e^bM^2}(-4M''M+2b'M'M+M'^2+4Me^b)=\frac{1}
{f_R}\left[\frac{e^{-b}(Rf_R-f)}{2}\right.\\\label{7a}&&
-\left.f_{R}'\left(\frac{a'e^a}{2e^b}\right)
+e^{a-b}f_{R}''+e^{a-b}f_{R}'\left(\frac{a'-b'}{2}+\frac{M'}{M}\right)
+e^a\rho_m\right],
\\\nonumber&&-\frac{1}{4M^2}(M'^2+2a'M'M-4Me^b)=\frac{1}{f_R}
\left[\frac{(f-Rf_R)}{2}-\frac{b'f_R'}{2}-f_{R}'\right.\\\label{8a}
&&\times\left.e^{a-b}f_{R}'\left(\frac{a'-b'}{2}+\frac{M'}{M}\right)
\left(\frac{a'-b'}{2}+\frac{M'}{M}\right)+e^bp_m\right],
\\\nonumber&&\frac{1}{4Me^b}(M'M(a'-b')+2M''M+M^2a'^2
-M^2a'b'-M'^2+2M^2a'')\\\nonumber&&=\frac{1}{f_R}\left[Mp_m+\frac{M'f_R'}{2e^bM}
+\frac{M(Rf_R-f)}{2}-\frac{f_R''}{Me^b}-\frac{f_R'}{Me^b}
\left(\frac{a'-b'}{2}+\frac{M'}{M}\right)\right].
\end{eqnarray}

The energy conditions provide a significant way to analyze physical
existence of some cosmological geometries. For WH geometry, the
violation of these conditions ensures the existence of a realistic
WH. To define energy conditions, Raychaudhari equations are
considered to be the most fundamental ingredients given as
\begin{eqnarray}\label{A}
\frac{d\theta}{d\tau}=-\frac{1}{3}\theta^2-\sigma_{\mu\nu}\sigma^{\mu\nu}
+\Theta_{\mu\nu}\Theta^{\mu\nu}-R_{\mu\nu}l^\mu l^\nu,\\\label{B}
\frac{d\theta}{d\tau}=-\frac{1}{2}\theta^2-\sigma_{\mu\nu}\sigma^{\mu\nu}
+\Theta_{\mu\nu}\Theta^{\mu\nu}-R_{\mu\nu}k^\mu k^\nu,
\end{eqnarray}
where $\theta,~l^\mu,~k^\mu,~\sigma$ and $\Theta$ represent
expansion scalar, timelike vector, null vector, shear and rotation
tensors. The first equation is defined for timelike congruence while
the second is for null congruence. The positivity of the last term
of both equations demands attractive gravity. For the
Einstein-Hilbert action, these conditions split into null (NEC)
($\rho_{m}+p_{m}\geq0$), weak (WEC)
($\rho_{m}\geq0,~\rho_{m}+p_{m}\geq0$), strong (SEC)
($\rho_{m}+p_{m}\geq0,~\rho_{m}+3p_{m}\geq0$) and dominant (DEC)
($\rho_{m}\geq0,~\rho_{m}\pm p_{m}\geq0$) energy conditions
\cite{I15}. As the Raychaudhari equations are found to be purely
geometric implying that $T^{(m)}_{\mu\nu}k^\mu k^\nu\geq0$ can be
replaced with $T^{eff}_{\mu\nu}k^\mu k^\nu\geq0$. Thus, the energy
conditions in $f(R)$ gravity turn out to be \cite{I16}
\begin{eqnarray*}\nonumber
\textbf{NEC}:\quad&&\rho_{eff}+p_{eff}\geq0,\\\nonumber
\textbf{WEC}:\quad&&\rho_{eff}\geq0,\quad\rho_{eff}+p_{eff}\geq0,\\\nonumber
\textbf{SEC}:\quad&&\rho_{eff}+p_{eff}\geq0,\quad\rho_{eff}+3p_{eff}\geq0,\\\nonumber
\textbf{DEC}:\quad&&\rho_{eff}\geq0,\quad\rho_{eff}\pm p_{eff}\geq0.
\end{eqnarray*}
Solving Eqs.(\ref{7a}) and (\ref{8a}), we obtain
\begin{eqnarray}\nonumber
p_m&=&-\frac{f}{2}+e^{-b}f_{R}'\left(\frac{a'}{2}+\frac{M'}{M}\right)
-\frac{f_R}{4e^bM^2}\left(2M'^2-4M''M-a'^2M^2\right.
\\\label{7}&+&\left.a'b'M^2+2b'M'M-2M^2a''\right),
\\\nonumber\rho_m&=&\frac{f_R}{4e^bM^2}\left(M^2a'^2-M^2a'b'
+2a'M'M+2M^2a''\right)+e^{-b}f_{R}''+e^{-b}f_{R}'\\\label{8}&\times&
\left(\frac{-b'}{2}+\frac{M'}{M}\right)+\frac{f}{2}.
\end{eqnarray}
In $f(R)$ gravity, NEC relative to the effective energy-momentum
tensor for (\ref{6}) yields
\begin{equation}\label{10}
\rho_{eff}+p_{eff}=\frac{1}{2e^{b}}\left(\frac{M'^2}{M^2}+\frac{a'M'}{M}
+\frac{b'M'}{M}-\frac{2M''}{M}\right).
\end{equation}

\section{Point-like Lagrangian}

In this section, we construct point-like Lagrangian corresponding to
the action (\ref{1}) via Lagrange multiplier approach. In this
regard, we consider following form of gravitational action
\cite{aop1}
\begin{equation}\label{C}
\mathcal{I}=\int\sqrt{-g}[f(R)-\lambda(R-\bar{R})]dr,
\end{equation}
where
\begin{eqnarray}\label{c2}
\sqrt{-g}&=&e^{\frac{a}{2}}e^{\frac{b}{2}}M,\quad\lambda=f_R,\\\nonumber
\bar{R}&=&\frac{1}{e^b}\left(-\frac{a'^2}{2}+\frac{a'b'}{2}-\frac{a'M'}{M}
-\frac{2M''}{M}+\frac{b'M'}{M}+\frac{M'^2}{2M^2}-a''+\frac{2e^b}{M}\right).
\end{eqnarray}
The dynamical constraint $\lambda$ is obtained by varying the action
(\ref{C}) with respect to $R$. In order to determine $p_m$, we
consider Bianchi identity ($\nabla_{\mu}T^{\mu\nu}$) whose radial
component gives
\begin{equation}\label{N1}
\frac{dp_m}{dr}+\frac{a'(r)}{2}\left(p_m+\rho_m\right)=0.
\end{equation}
Solving this differential equation with $p_m=\omega\rho_m$, it
follows that
\begin{equation}\label{c1}
\rho_m=\rho_0a^{-\frac{(1+\omega)}{2\omega}},\quad
p_m=\omega\rho_m=\omega\rho_0a^{-\frac{(1+\omega)}{2\omega}},
\end{equation}
where $\omega$ represents equation of state parameter. Inserting
Eq.(\ref{c2}) and (\ref{c1}) in (\ref{C}), we obtain
\begin{eqnarray}\nonumber
\mathcal{I}&=&\int
e^{\frac{a-b}{2}}M\left[f(R)-Rf_R+\frac{f_R}{e^b}\left(-\frac{a'^2}{2}
+\frac{a'b'}{2}-\frac{a'M'}{M}-\frac{2M''}{M}+\frac{b'M'}{M}
\right.\right.\\\label{C1}&+&\left.\left.\frac{M'^2}{2M^2}-a''
+\frac{2e^b}{M}\right)+\omega\rho_0a^{-\frac{(1+\omega)}{2\omega}}\right]dr.
\end{eqnarray}
Eliminating second order derivatives via integration by parts from
the above action and following Lagrangian density definition, we
obtain point-like Lagrangian as
\begin{eqnarray}\nonumber
&&\mathcal{L}(r,a,b,M,R,a',b',M',R')=e^{\frac{a}{2}}e^{\frac{b}{2}}M\left(f-Rf_R
+\omega\rho_0a^{-\frac{(1+\omega)}{2\omega}}+\frac{2f_R}{M}\right)
\\\label{11}&&+\frac{e^{\frac{a}{2}}M}{e^{\frac{b}{2}}}\left\{f_R\left(\frac{M'^2}
{2M^2}+\frac{a'M'}{M}\right)+f_{RR}\left(a'R'+\frac{2M'R'}{M}\right)\right\}.
\end{eqnarray}
For static spherically symmetric spacetime, the Euler-Lagrange
equation and Hamiltonian of the dynamical system or energy function
associated with point-like Lagrangian are defined as
\begin{eqnarray}\nonumber
&&\frac{\partial\mathcal{L}}{\partial
q^i}-\frac{dp_i}{dr}=0,\quad\mathcal{H}=\sum_iq'^{i}p_i-\mathcal{L},
\end{eqnarray}
where $q^i$ are generalized coordinates and
$p_i=\frac{\partial\mathcal{L}}{\partial{q'^i}}$ represents
conjugate momenta. The variation of Lagrangian with respect to
configuration space leads to
\begin{eqnarray*}
&&e^b\left(f-Rf_R
+\omega\rho_0a^{-\frac{(1+\omega)}{2\omega}}-(1+\omega)\rho_0
a^{-\frac{(1+3\omega)}{2\omega}}+\frac{2f_R}{M}\right)+\left(\frac{M'^2}
{2M^2}+\frac{b'M'}{M}\right.\\\nonumber&&-\left.\frac{2M''}
{M}\right)f_R+f_{RR}\left(b'R'-2R''
-\frac{2M'R'}{M}\right)-2R'^2f_{RRR}=0,
\\\nonumber&&e^b\left(f-Rf_R+\omega\rho_0a^{-\frac{(1+\omega)}
{2\omega}}+\frac{2f_R}{M}\right)-f_R\left(\frac{M'^2}
{2M^2}+\frac{a'M'}{M}\right)-f_{RR}\left(a'R'\right.
\\\nonumber&&+\left.\frac{2M'R'}{M}\right)=0,
\\\nonumber&&e^b\left(f-Rf_R+\omega\rho_0a^{-\frac{(1+\omega)}
{2\omega}}+\frac{2f_R}{M}\right)
+f_R\left(-\frac{a'^2}{2}+\frac{a'b'}{2}-\frac{a'M'}{2M}
-\frac{M''}{M}-a''\right.\\\nonumber&&+\left.\frac{b'M'}
{2M}+\frac{M'^2}{2M^2}\right)+f_{RR}
\left(b'R'-a'R'-2R''-\frac{M'R'}{M}\right)-2R'^2f_{RRR}=0,
\\\nonumber&&\left[e^b\left(\frac{2}{M}-R\right)
-\frac{a'^2}{2}+\frac{a'b'}{2}-\frac{a'M'}{M}
-\frac{2M''}{M}+\frac{b'M'}{M}+\frac{M'^2}{2M^2}-a''\right]f_{RR}=0.
\end{eqnarray*}
The energy function and variation of Lagrangian relative to shape
function yield
\begin{equation}\label{12}
e^b=\frac{\frac{f_R{M}'}{{M}}\left(\frac{{M}'}{2{M}^2}+{a}'{M}'\right)
+R'f_{RR}({a}'{M}+2{M}')}{f-Rf_R
+\omega\rho_0{a}^{-\frac{(1+\omega)}{2\omega}}+\frac{2f_R}{{M}}}.
\end{equation}

\section{Noether Symmetry Approach}

The physical characteristics of a dynamical system can be identified
by constructing the associated Lagrangian which successfully
describes energy content and the existence of possible symmetries of
the system. In this regard, Noether symmetry approach provides an
interesting way to construct new cosmological models and geometries
in modified theories of gravity. According to well-known Noether
theorem, group generator yields associated conserved quantity if
point-like Lagrangian remains invariant under a continuous group. In
order to investigate the presence of Noether symmetry and relative
conserved quantity of static spherically symmetric metric, we
consider a vector field \cite{aop2}
\begin{eqnarray}\label{13}
K&=&\tau(r,q^i)\frac{\partial}{\partial
r}+\zeta^i(r,q^i)\frac{\partial}{\partial q^i},
\end{eqnarray}
where $r$ behaves as an affine parameter while $\tau$ and $\zeta^i$
are unknown coefficients of the vector field $K$.

The presence of Noether symmetry is assured only if point-like
Lagrangian satisfies the invariance condition and the vector field
is found to be unique on tangent space. Consequently, the vector
field acts as a symmetry generator generating associated conserved
quantity. In this case, the invariance condition is defined as
\begin{equation}\label{14}
K^{[1]}\mathcal{L}+(D\tau)\mathcal{L}=DB(r,q^i),
\end{equation}
where $B$ denotes boundary term of the extended symmetry, $K^{[1]}$
describes first order prolongation and $D$ represents total
derivative given by
\begin{equation}\label{15}
K^{[1]}=K+(D\zeta^i-{q'}^iD\tau)\frac{\partial}{\partial
{q'}^i},\quad D=\frac{\partial}{\partial
r}+{q'}^{i}\frac{\partial}{\partial q^i}.
\end{equation}
Noether symmetries coming from invariance condition (\ref{14}) lead
to identify associated conserved quantities through first integral.
If the Lagrangian remains invariant under translation in time and
position, then the first integral identifies energy and linear
momentum conservation while rotationally symmetric Lagrangian yields
conservation of angular momentum \cite{13}. For invariance condition
(\ref{14}), the first integral is defined as
\begin{equation}\label{16}
\Sigma=B-\tau\mathcal{L}-(\zeta^i-{q'}^i\tau)
\frac{\partial\mathcal{L}}{\partial {q'}^i}.
\end{equation}

For configuration space $Q=\{a,b,M,R\}$, the vector field $K$ and
first order prolongation $K^{[1]}$ take the following form
\begin{eqnarray}\nonumber
K&=&\tau\frac{\partial}{\partial r}+\alpha\frac{\partial}{\partial
a}+\beta\frac{\partial}{\partial b}+\gamma\frac{\partial}{\partial
M}+\delta\frac{\partial}{\partial R},\quad
K^{[1]}=\tau\frac{\partial}{\partial
r}+\alpha\frac{\partial}{\partial a}+\beta\frac{\partial}{\partial
b}\\\label{17}&+&\gamma\frac{\partial}{\partial
M}+\delta\frac{\partial}{\partial R}+\alpha'\frac{\partial}{\partial
a'}+\beta'\frac{\partial}{\partial
b'}+\gamma'\frac{\partial}{\partial
M'}+\delta'\frac{\partial}{\partial R'},
\end{eqnarray}
where the radial derivative of unknown coefficients of vector field
are defined as
\begin{eqnarray}\label{18}
\sigma'_{_j}&=&D\sigma_{_j}-{q'}^iD\tau,\quad j=1...4.
\end{eqnarray}
Here $\sigma_1,~\sigma_2,~\sigma_3$ and $\sigma_4$ correspond to
$\alpha,~\beta,~\gamma$ and $\delta$, respectively. Inserting
Eqs.(\ref{11}), (\ref{17}) and (\ref{18}) in (\ref{14}) and
comparing the coefficients of $a'^2,~a'b'M',~a'M'^2$ and $a'R'^2$,
we obtain
\begin{equation}\label{19}
\tau,_{_a}f_R=0,\quad\tau,_{_b}f_R=0,\quad\tau,_{_M}f_R=0,\quad\tau,_{_R}f_{RR}=0.
\end{equation}
This equation implies that either $f_R=0$ or vice verse. The first
choice leads to trivial solution. Therefore, we consider $f_R\neq0$
and compare the remaining coefficients which yield the following
system of equations
\begin{eqnarray}\label{20}
&&B,_{_b}=0,\quad\tau,_{_a}=0,\quad\tau,_{_b}=0,\quad\tau,_{_M}=0,\quad\tau,_{_R}=0,
\\\label{21}&&e^{\frac{a}{2}}(\gamma,_{_r}f_R+M
\delta,_{_r}f_{RR})=e^{\frac{b}{2}}B,_{_a},\\\label{22a}&&e^{\frac{a}{2}}
(\alpha,_{_r}M+2\gamma,_{_r})f_{RR}=e^{\frac{b}{2}}B,_{_R},\\\label{22}&&e^{\frac{a}{2}}
(\alpha,_{_r}f_R+\gamma,_{_r}M^{-1}f_R+2\delta,_{_r}f_{RR})=e^{\frac{b}{2}}B,_{_M},
\\\label{23}&&\gamma,_{_a}f_R+M\delta,_{_a}f_{RR}=0,\\\label{24}&&\gamma,_{_a}
f_R+M\delta,_{_a}f_{RR}=0,\\\label{25}&&\alpha,_{_b}f_R+\gamma,_{_b}M^{-1}
f_R+2\delta,_{_b}f_{RR}=0,\\\label{26}&&M\alpha,_{_b}f_{RR}+2\gamma,_{_b}f_{RR}=0,
\\\label{27}&&M\alpha,_{_R}f_{RR}+2\gamma,_{_R}f_{RR}=0,\\\label{28}&&f_R
(\alpha-\beta-2\gamma
M^{-1}+4M\alpha,_{_M}+4\gamma,_{_M}-2\tau,_{_r})+f_{RR}(2\delta+8M\delta,_{_M})=0,
\\\label{29}&&f_R(\alpha-\beta+2\alpha,_{_a}-2\tau,_{_r}
+2\gamma,_{_M}+2\gamma,_{_a})+f_{RR}(2\delta+2M\delta,_{_M}+4\delta,_{_a})=0,
\\\nonumber&&f_R(\alpha,_{_R}+\gamma,_{_R}M^{-1})+f_{RR}(\alpha-\beta+M\alpha,_{_M}
+2\gamma,_{_M}-2\tau,_{_r}+2\delta,_{_R})+2\delta\\\label{30}&&\times
f_{RRR}=0,\\\nonumber&&2\gamma,_{_R}f_R+f_{RR}(M\alpha-M\beta+2\gamma
+2M\alpha,_{_a}-2M\tau,_{_r}+4\gamma,_{_a}+2M\delta,_{_R})+2M\\\label{31}&&\times\delta
f_{RRR}=0,\\\nonumber&&e^{\frac{a}{2}}e^{\frac{b}{2}}M\{\frac{1}{2}(f-Rf_R
+\omega\rho_0a^{-\frac{(1+\omega)}{2\omega}}+\frac{2f_R}{M})(\alpha+\beta+\tau,_{_r})
-\frac{1}{2}\alpha(1+\omega)\rho_0\\\nonumber&&\times
a^{-\frac{(1+3\omega)}{2\omega}}+\delta
M(2M^{-1}-R)f_{RR}\}+e^{\frac{a}{2}}e^{-\frac{b}{2}}\gamma(f-Rf_R
+\omega\rho_0a^{-\frac{(1+\omega)}{2\omega}})\\\label{32}&&=B,_{_r}.
\end{eqnarray}

In order to solve this system, we consider $M(r)=r^2$ and taking
$B,_{_a},~B,_{_M},~B,_{_R}=0$, Eqs.(\ref{20})-(\ref{27}) give
\begin{equation*}
\alpha=Y_2(a,r),\quad\gamma=Y_1(r),\quad\delta=Y_3(r,R).
\end{equation*}
Inserting these values in Eqs.(\ref{28})-(\ref{31}), we obtain
\begin{equation*}
Y_1(r)=0,\quad Y_2(a,r)=c_2,\quad
Y_3(r,R)=\frac{c_1f_R}{f_{RR}},\quad\beta=2c_1+c_2-2\tau,_{_r},
\end{equation*}
where $c_1$ and $c_2$ are arbitrary constants. For these solutions,
the coefficients of symmetry generator turn out to be
\begin{equation}
\alpha=c_2,\quad\beta=2c_1+c_2,\quad\gamma=0,
\quad\delta=\frac{c_1f_R}{f_{RR}},\quad\tau=c_0.
\end{equation}
Substituting these coefficients in Eq.(\ref{32}), we formulate
boundary term and explicit form of $f(R)$ as follows
\begin{eqnarray*}
f(R)&=&-\frac{1}{2(c_1+c_2)}\left[-(1+\omega)\rho_0a^{-\frac{(1
+3\omega)}{2\omega}}+2\omega(c_1+c_2)\rho_0a^{-\frac{(1+\omega)}
{2\omega}}\right.\\\nonumber
&-&\left.6c_4e^{\frac{-a-b}{2}}\right],\quad B=c_3+c_4r^3.
\end{eqnarray*}
The coefficients of symmetry generator, boundary term and solution
of $f(R)$ satisfy the system of Eqs.(\ref{20})-(\ref{31}) for
$c_1=0$. Thus, the symmetry generator and the corresponding first
integral take the form
\begin{eqnarray*}
K&=&c_0\frac{\partial}{\partial r}+c_2\frac{\partial}{\partial
a}+c_2\frac{\partial}{\partial
b},\\\nonumber\Sigma&=&c_3+c_4r^3-c_0\left[e^{\frac{a}{2}}
e^{\frac{b}{2}}r^2(f-Rf_R
+\omega\rho_0a^{-\frac{(1+\omega)}{2\omega}}+2f_Rr^{-2})
\right.\\\nonumber&+&\left.\frac{e^{\frac{a}{2}}r^2}
{e^{\frac{b}{2}}}\{f_R(2r^{-2}+2a'r^{-1})
+f_{RR}(a'R'+4R'r^{-1})\}\right]\\\nonumber&-&c_2
e^{\frac{a-b}{2}}(R'r^2f_{RR}+2rf_R).
\end{eqnarray*}

The verification of Eq.(\ref{32}) yields
\begin{equation}\label{33}
b(r)=\int\frac{8c_6r^2+a''r^2+4a'r'+a'^2r^2-4c_7}{r(4+a'r)}dr+c_5,
\end{equation}
where $c_i$'s $(i=3...8)$ are arbitrary constants and this solution
satisfies Eq.(\ref{32}) for $\omega=1,1/3,-1/3,-1$. To discuss
physical features and geometry of WH via shape function, we take
red-shift function, $a(r)=k$ and $a(r)=-\frac{k}{r},~k>0$, where $k$
denotes constant \cite{I17}. In the following, we solve integral for
both choices of red-shift function.

\subsubsection*{Case I: $a(r)=k$}

We first consider red-shift function to be constant and evaluate
$b(r)$ such as
\begin{equation}\label{34}
b(r)=c_6r^2-c_7\ln r+c_5.
\end{equation}
Consequently, the shape function turns out to be
\begin{equation}\label{35}
h(r)=r(1-e^{-b(r)})=r(1-c_7re^{-c_6r^2-c_5}).
\end{equation}
In this case, the explicit form of $f(R)$ reduces to
\begin{eqnarray}\nonumber
f(R)=-\frac{1}{2c_2}\left[-(1+\omega)\rho_0
k^{-\frac{(1+3\omega)}{2\omega}}+2\omega
c_2\rho_0k^{-\frac{(1+\omega)}{2\omega}}
-6c_4\sqrt{c_7r}e^{\frac{-c_6r^2-c_5-k}{2}}\right].\\\label{A1}
\end{eqnarray}

The $f(R)$ theory of gravity is one of the competitive candidates in
modified theories of gravity as it naturally unifies two expansion
phases of the universe, i.e., inflation at early times and cosmic
acceleration at current epoch. The higher derivative of curvature
terms with positive power are dominant at the early universe leading
to the inflationary stage. The terms with negative power of the
curvature serve as gravitational alternative for the dark energy
that acts as a possible source to speed-up cosmic expansion
\cite{aop1a}. Despite the fact that the ghost-free $f(R)$ theory is
very interesting and useful as it passes solar system tests, it also
suffers from instabilities. For instance, the theory with
$\frac{1}{R}$ may develop the instability \cite{aop22a} whereas by
adding a term of $R^2$ to this specific form of $f(R)$ model, one
can easily eliminate this instability \cite{aop9a}. Therefore, the
viable $f(R)$ models require to satisfy the following stability
constraints $f_R(R)>0,~f_{RR}(R)>0,~R>R_0$ where $R_0$ is the
current Ricci scalar \cite{aop25a}.

In Figure \textbf{1}, both plots indicate that the constructed
$f(R)$ model (\ref{A1}) preserves the stability conditions. Figure
\textbf{2} shows the graphical analysis of shape function. The upper
left plot represents positive behavior of $h(r)$ while the upper
right indicates that the shape function admits asymptotic behavior.
The lower left plot locates the WH throat at $r_0=4.4$ and the
corresponding right plot identifies that
$\frac{dh(r_0)}{dr}=0.9427<1$. To discuss physical existence of WH,
we insert constant red-shift function and Eq.(\ref{34}) in
(\ref{10}) yielding
\begin{equation*}
\rho_{eff}+p_{eff}=\frac{rh'(r)-h(r)}{r^3}<0,
\end{equation*}
which satisfies the flaring-out condition. Consequently, NEC
violates in this case, $\rho_{eff}+p_{eff}<0$ and assures the
presence of repulsive gravity leading to traversable WH. In order to
study the realistic existence of traversable WH, we analyze the
behavior of NEC and WEC in Figure \textbf{3}. Both plots indicate
that energy density and pressure recover energy bounds as
$\rho_m\geq0$ and $\rho_m+p_m\geq0$ implying physically acceptable
traversable WH.
\begin{figure}\centering{\epsfig{file=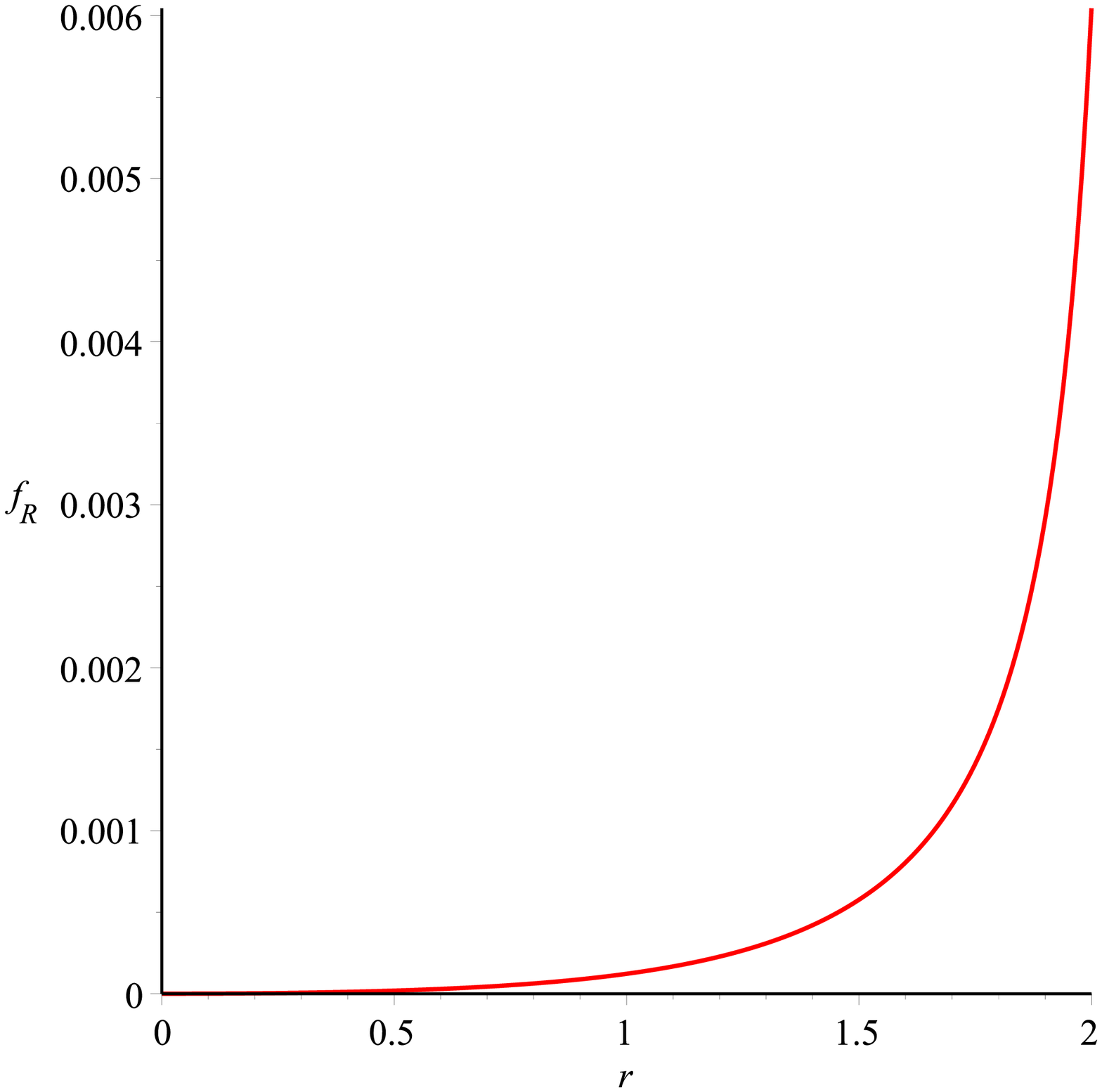,
width=0.4\linewidth}\epsfig{file=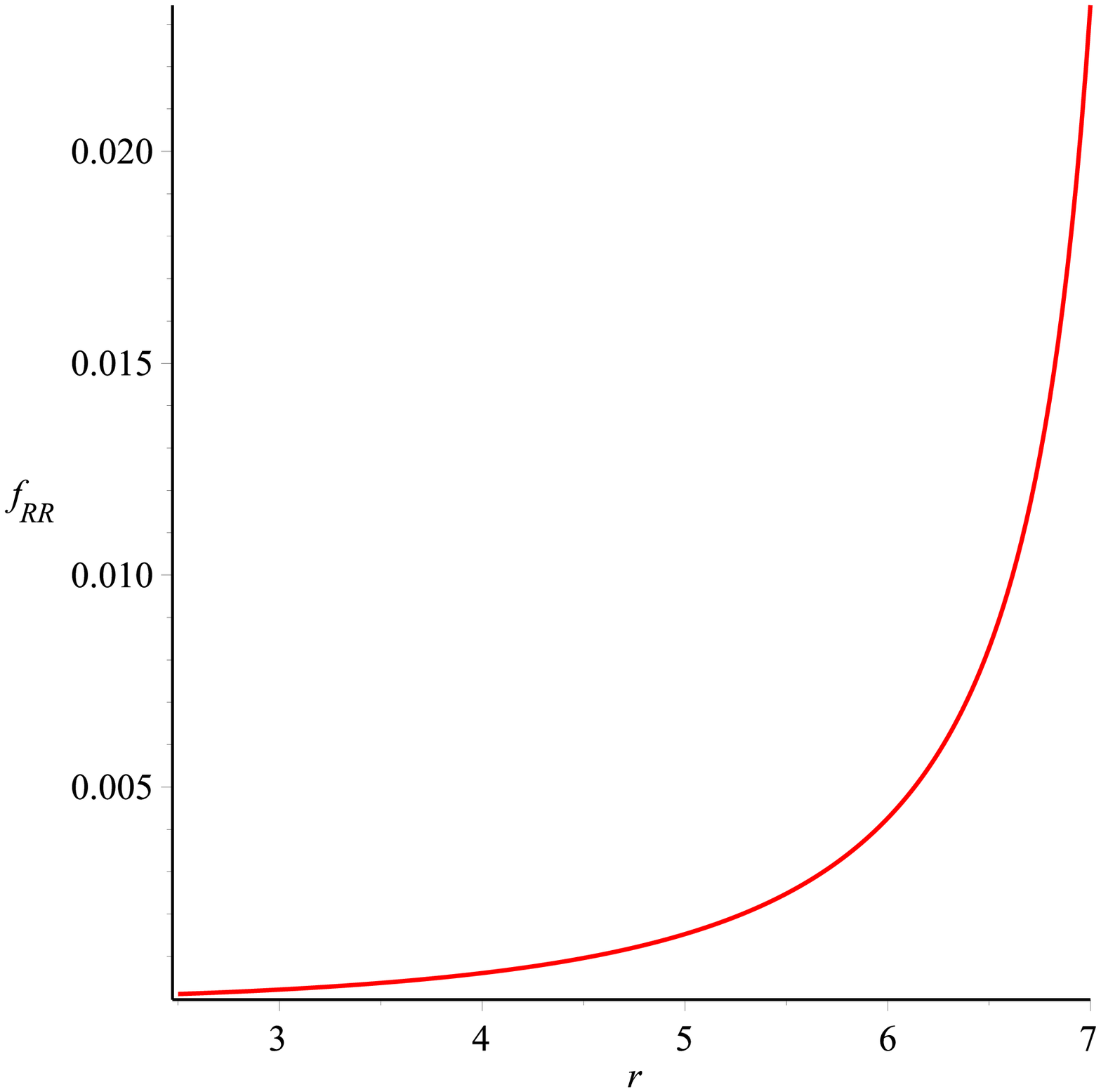,
width=0.4\linewidth}\caption{Plots of stability conditions of $f(R)$
model versus $r$ for $c_{_2}=5$, $c_{_4}=0.01$, $c_{_5}=-0.35$,
$c_{_6}=0.1$, $c_{_7}=-0.25$, $\rho_0=1$ and $k=0.5$.}}
\end{figure}
\begin{figure}\centering{
\epsfig{file=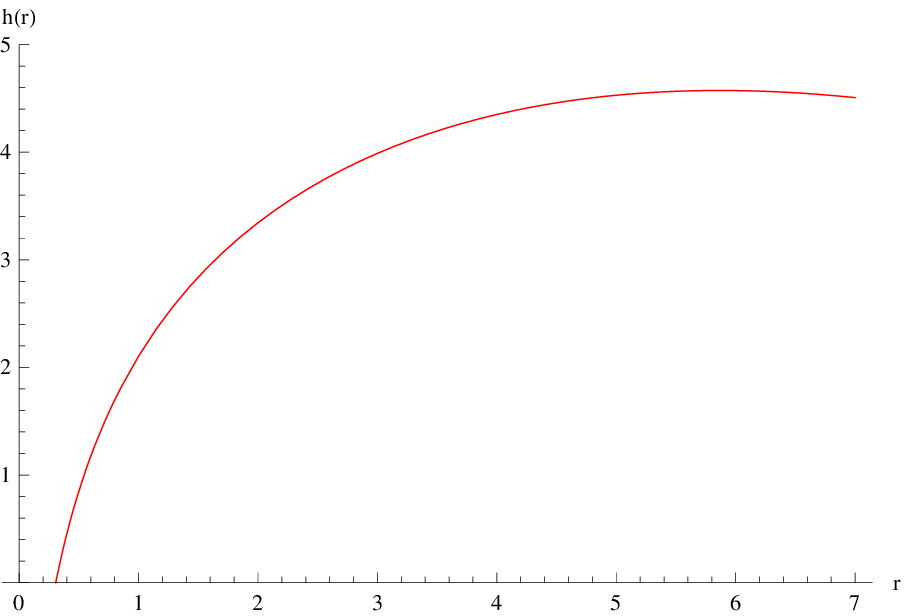, width=0.4\linewidth}\epsfig{file=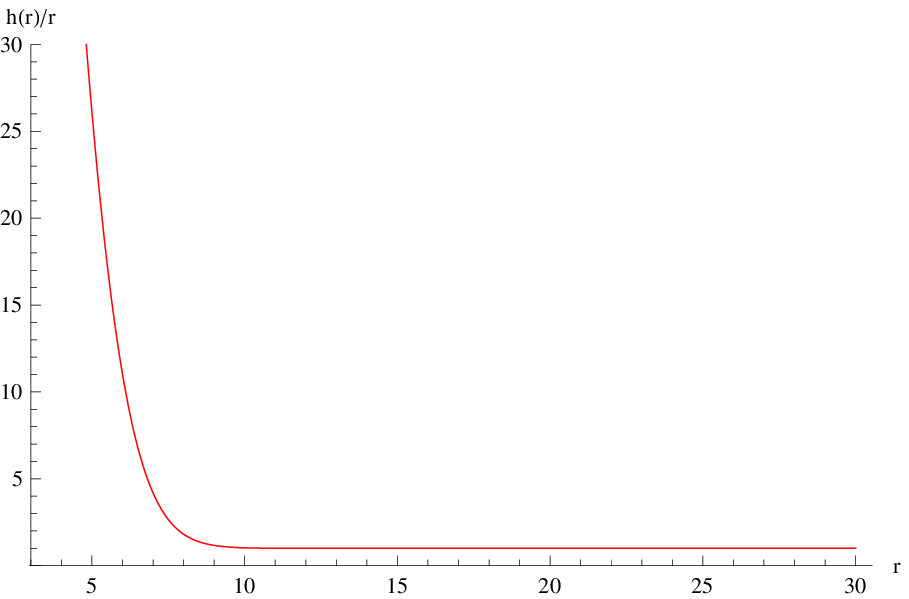, width=0.4\linewidth}\\
\epsfig{file=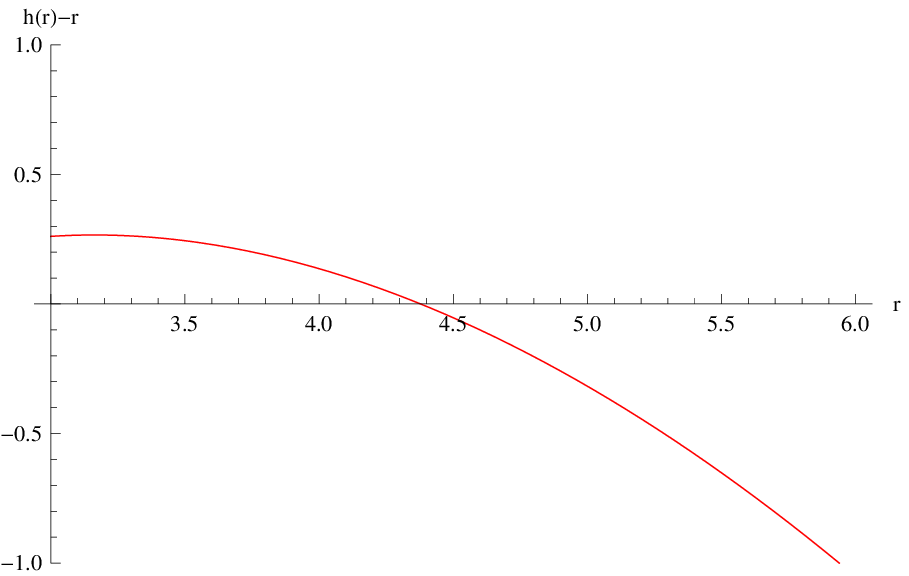, width=0.4\linewidth}\epsfig{file=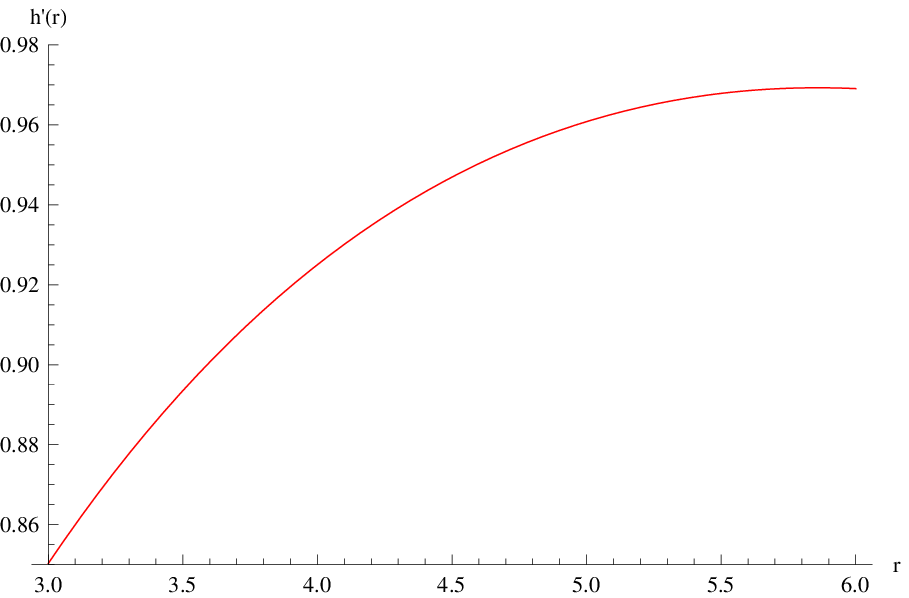,
width=0.4\linewidth} \caption{Plots of $h(r),~\frac{h(r)}
{r},~h(r)-r$ and $\frac{dh(r)}{dr}$ versus $r$ for $c_{_5}=-0.35$,
$c_{_6}=0.1$ and $c_{_7}=-0.25$.}}
\end{figure}
\begin{figure}\centering{
\epsfig{file=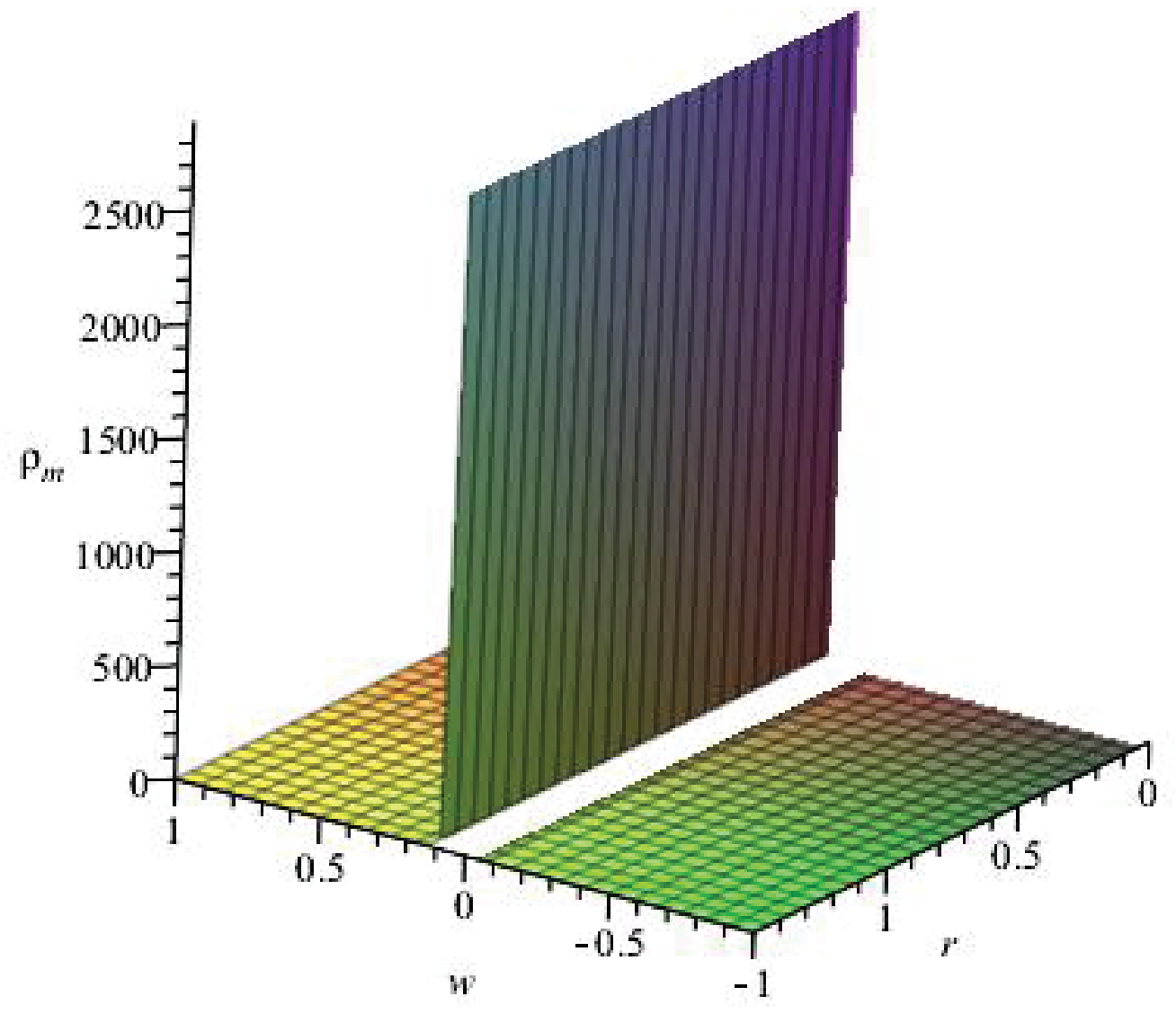, width=0.5\linewidth}\epsfig{file=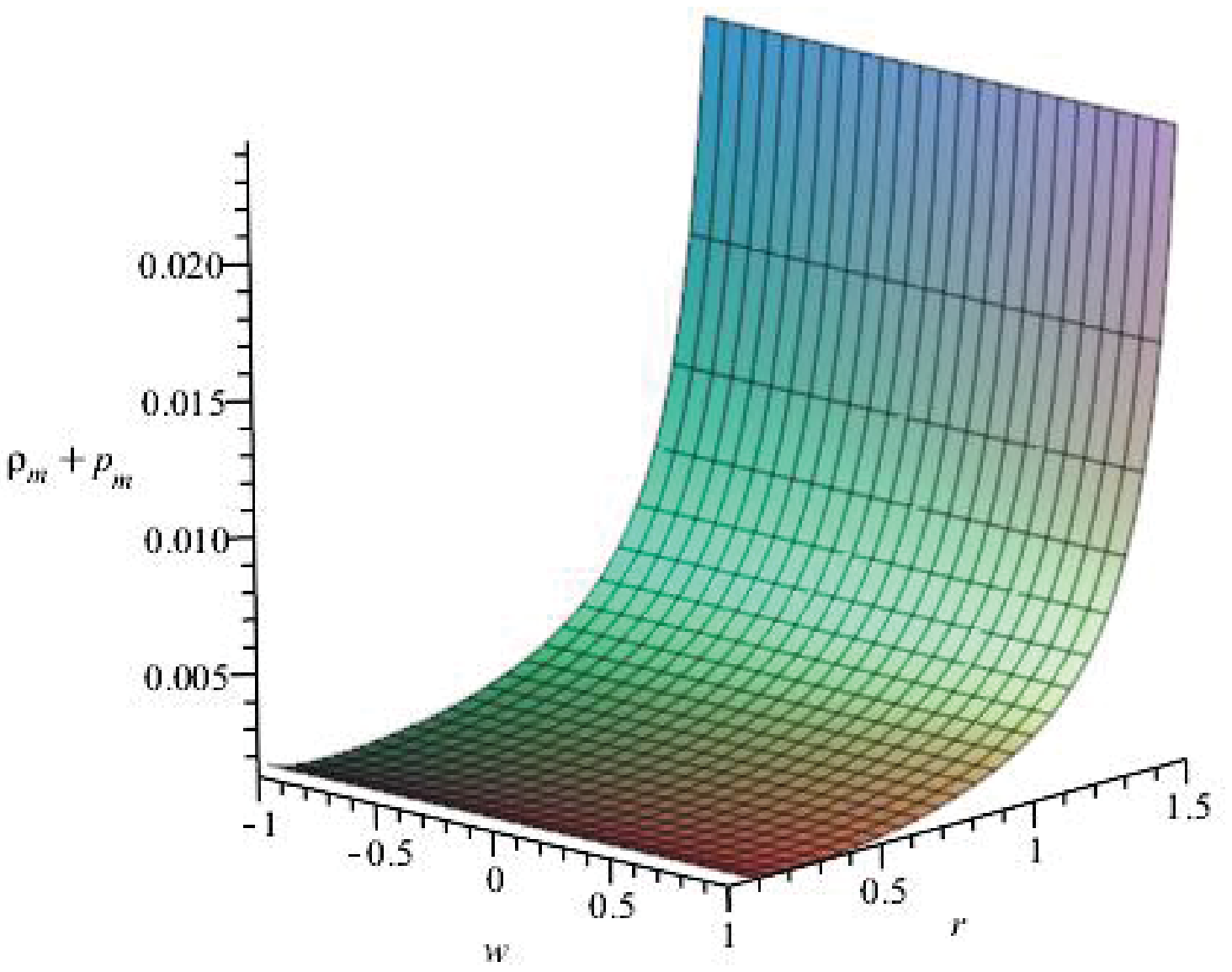,
width=0.5\linewidth}\caption{Plots of $\rho_m$ and $\rho_m+p_m$
versus $r$.}}
\end{figure}

\subsubsection*{case II: $a(r)=-k/r$}

In this case, we choose red-shift function in terms of $r$ leading
to
\begin{eqnarray}\nonumber
a(r)&=&-\frac{k}{r},\quad b(r)=\frac{1}{8r}(4c_6r^2(2r-k)-32c_8r\ln
r+(32r-8c_7r+c_6kr^2)\\\label{36}&\times&\ln(4r+k)-8k/c_8)+c_5,\quad
k>0.
\end{eqnarray}
For this solution of $a(r)$ and $b(r)$, the generic function takes
the form
\begin{eqnarray}\nonumber
f(R)&=&-\frac{1}{2c_2}\left[-(1+\omega)\rho_0\left(-\frac{k}
{r}\right)^{-\frac{(1+3\omega)}{2\omega}}+2\omega
c_2\rho_0\left(-\frac{k}{r}\right)^{-\frac{(1+\omega)}{2\omega}}-6c_4
\right.\\\label{A2}&\times&\left.\sqrt{c_8r^4(4r+k)^{-4+c_7-\frac{k^2c_6}
{8}}}e^{\frac{-(c_6r^2-\frac{c_6kr}{2}-\frac{k}{c_8r})-c_5+k}{2}}\right].
\end{eqnarray}
The corresponding shape function becomes
\begin{eqnarray}\label{37}
h(r)=r(1-c_8r^4(4r+k)^{-4+c_7-\frac{k^2c_6}{8}}e^{-(c_6r^2-\frac{c_6kr}{2}
-\frac{k}{c_8r})-c_5}).
\end{eqnarray}
\begin{figure}\centering{\epsfig{file=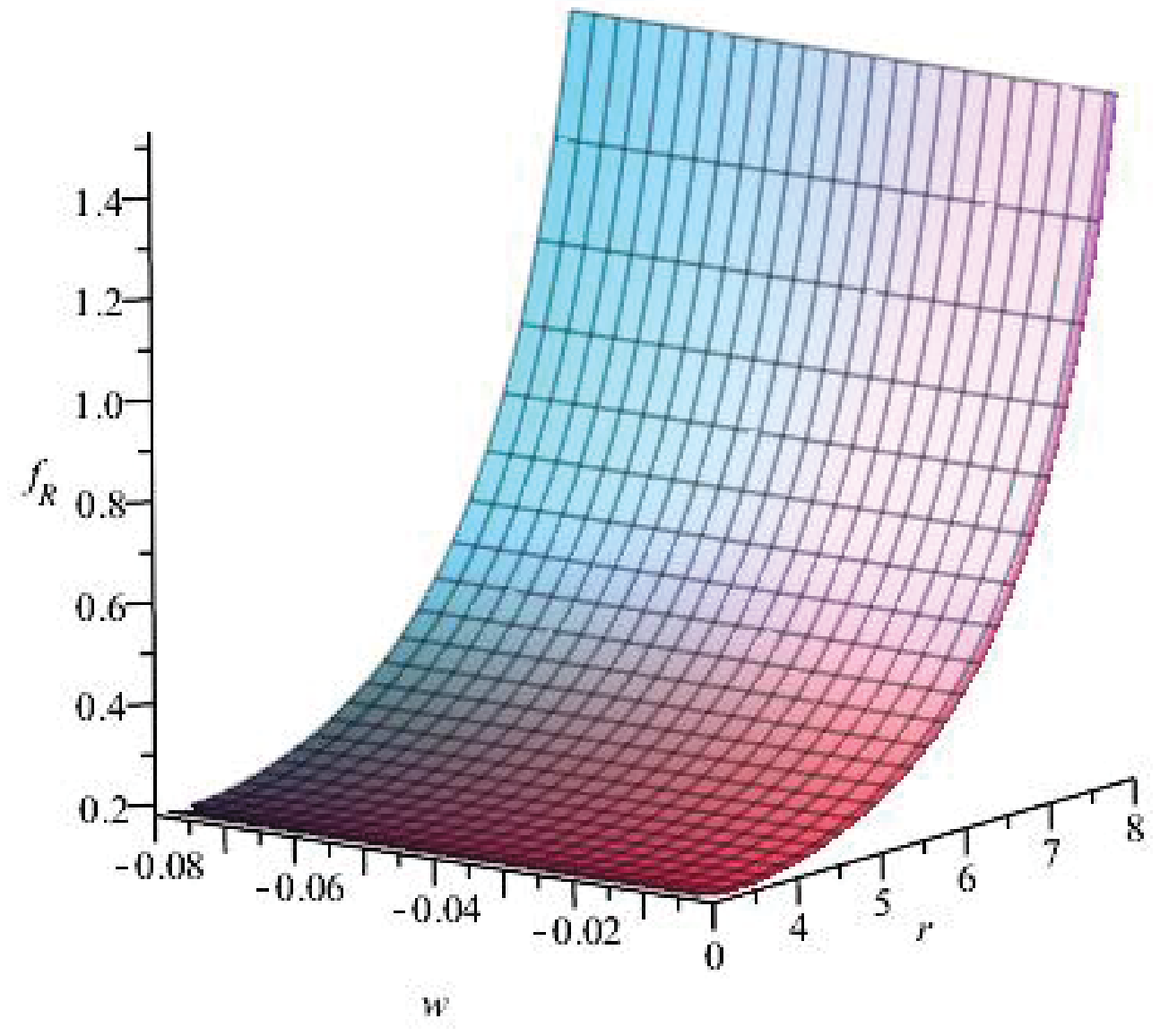,
width=0.55\linewidth}\epsfig{file=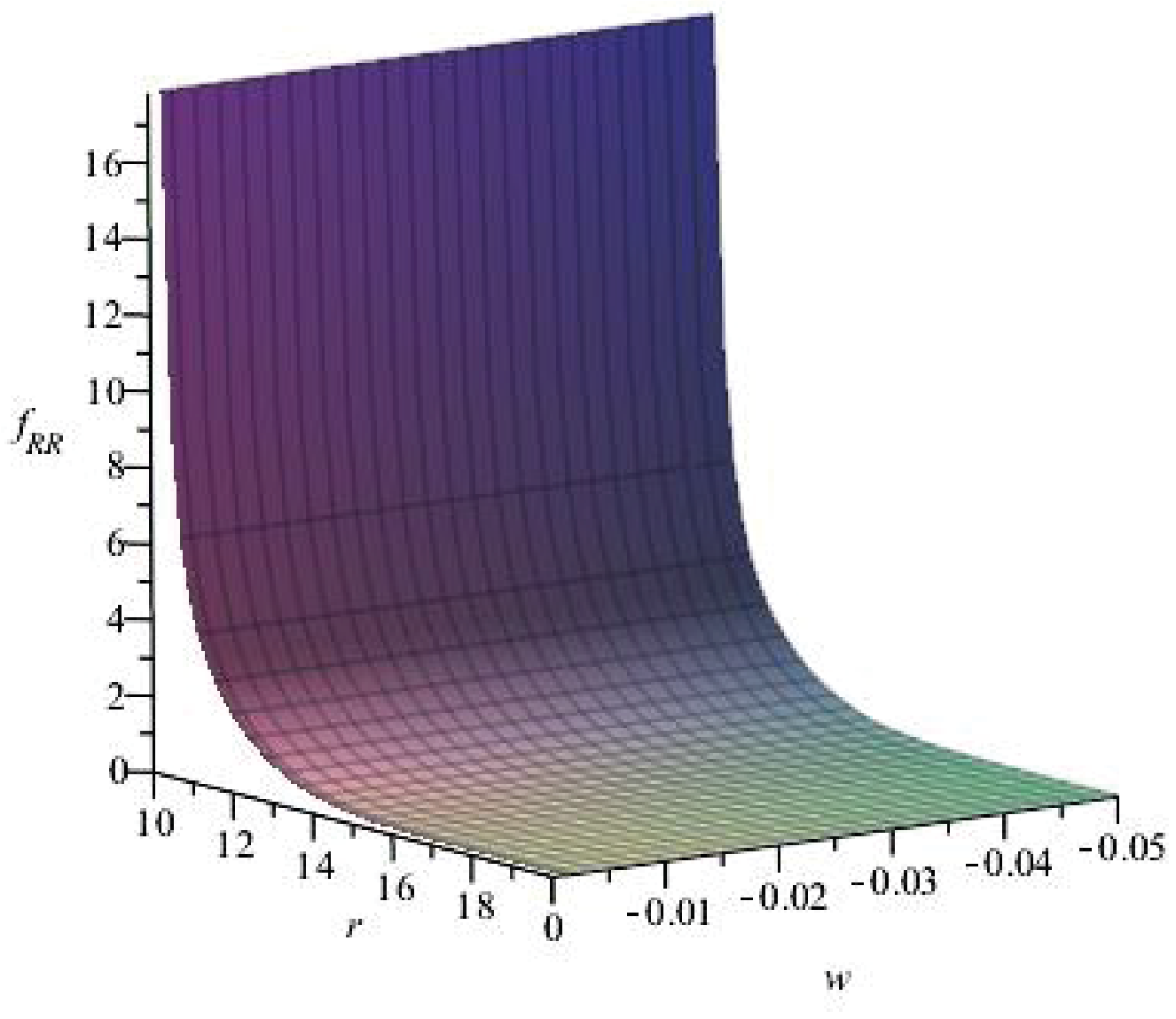,
width=0.45\linewidth}\caption{Stability conditions of $f(R)$ versus
$r$ for $c_{_2}=5$, $c_{_4}=0.01$, $c_{_5}=-0.35$, $c_{_6}=0.1$,
$c_{_7}=-0.25$ and $k=0.5$.}}
\end{figure}
\begin{figure}\centering{
\epsfig{file=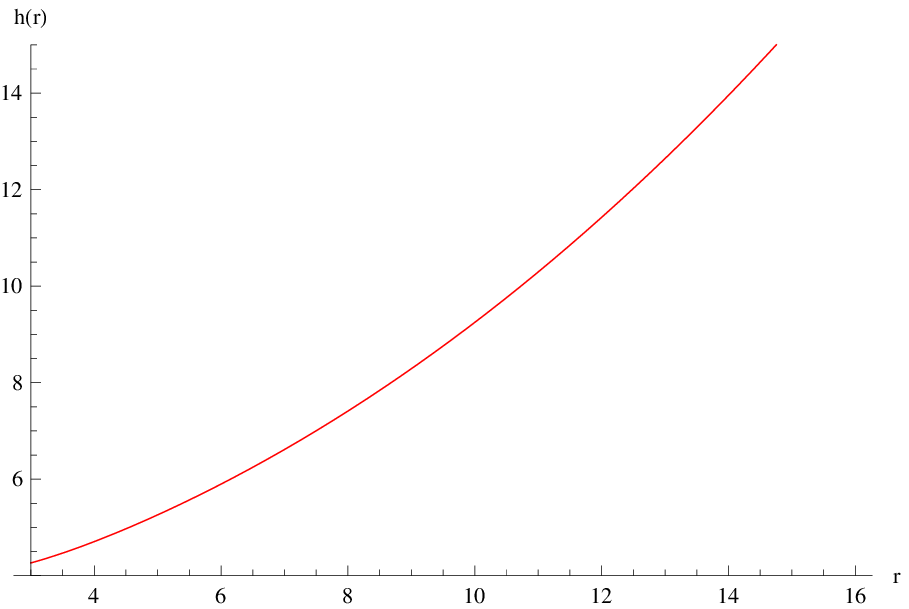, width=0.4\linewidth}\epsfig{file=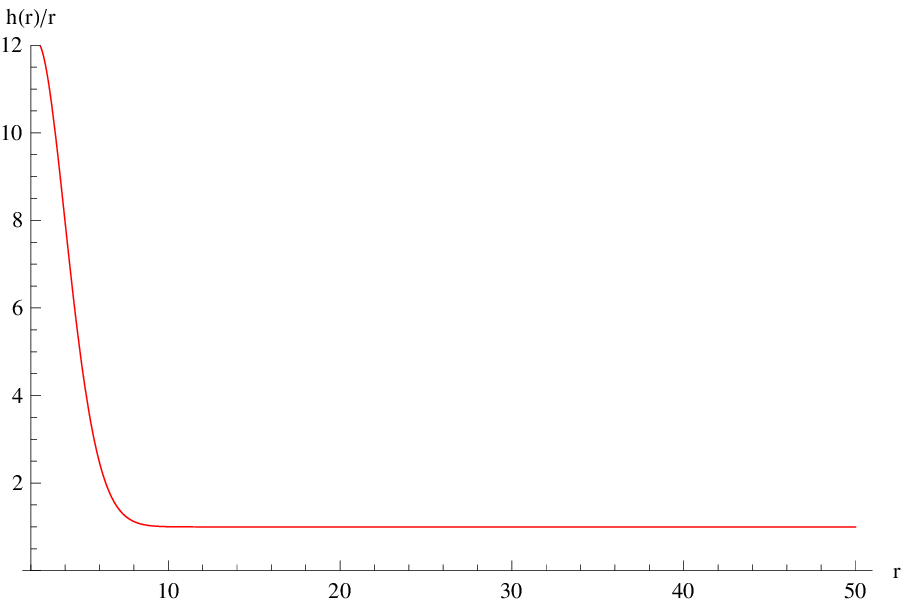, width=0.4\linewidth}\\
\epsfig{file=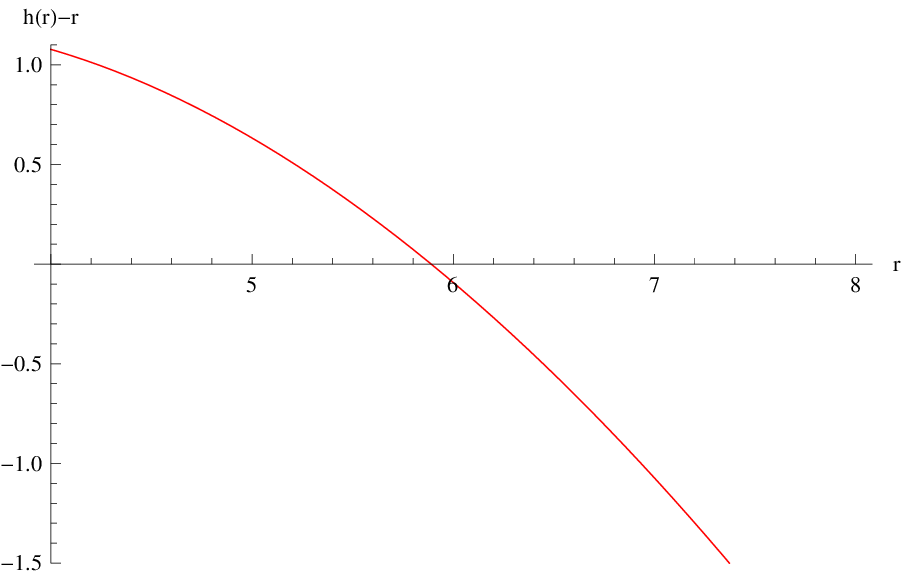, width=0.4\linewidth}\epsfig{file=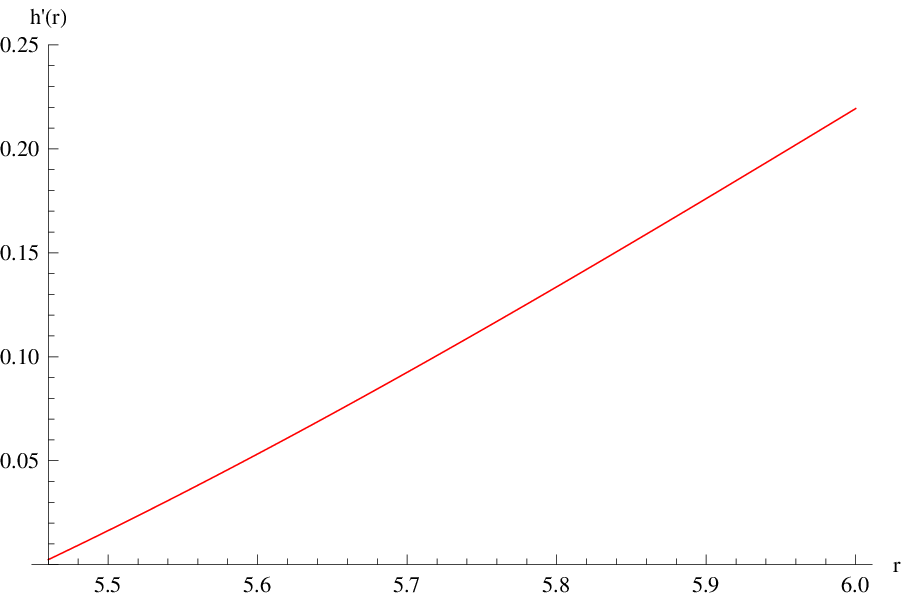,
width=0.4\linewidth} \caption{Plots of $h(r),~\frac{h(r)}
{r},~h(r)-r$ and $\frac{dh(r)}{dr}$ versus $r$ for $c_{_5}=-4$,
$c_{_6}=0.1$, $c_{_8}=-1$ and $k=0.25$.}}
\end{figure}
\begin{figure}\centering\epsfig{file=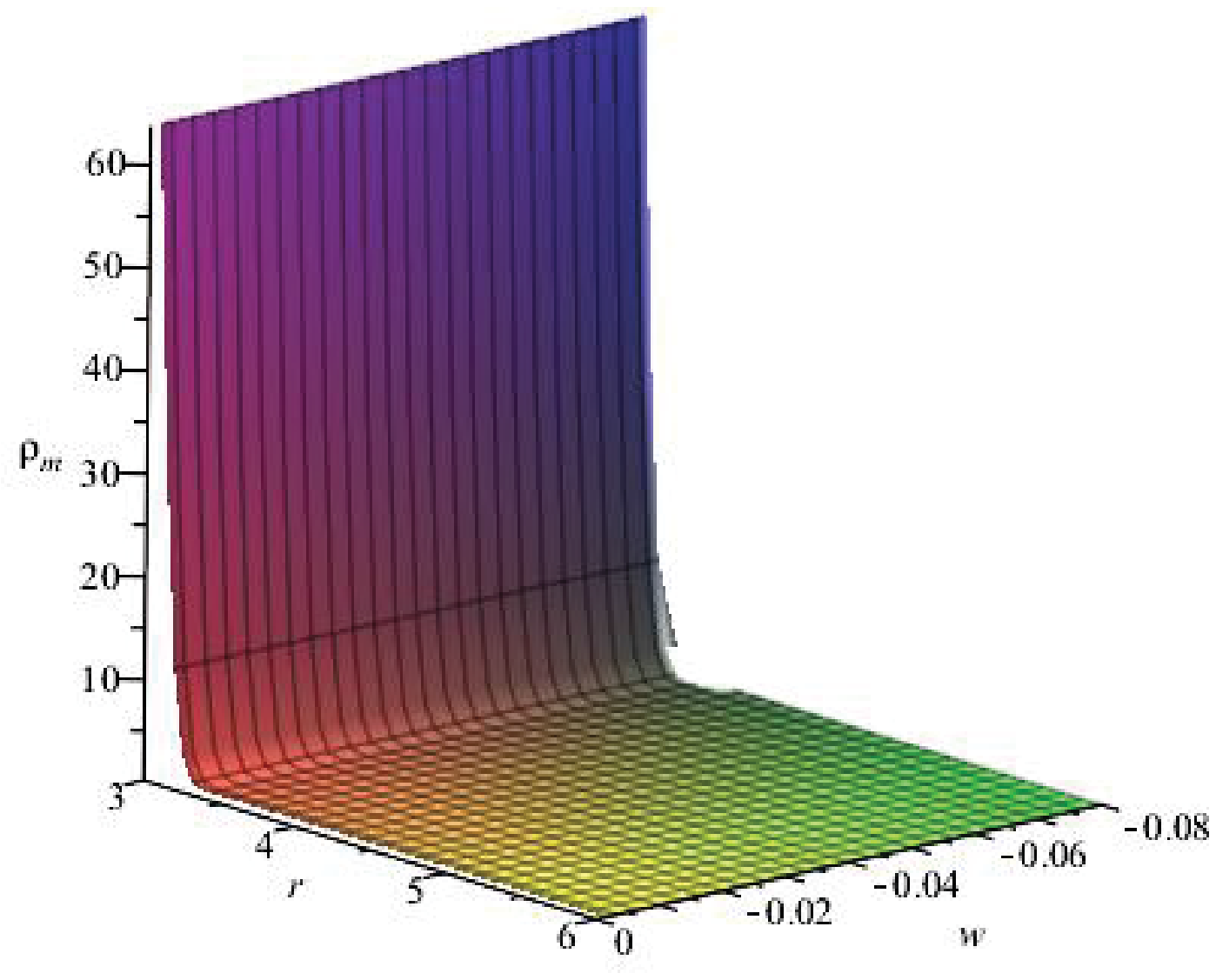,
width=0.5\linewidth}\epsfig{file=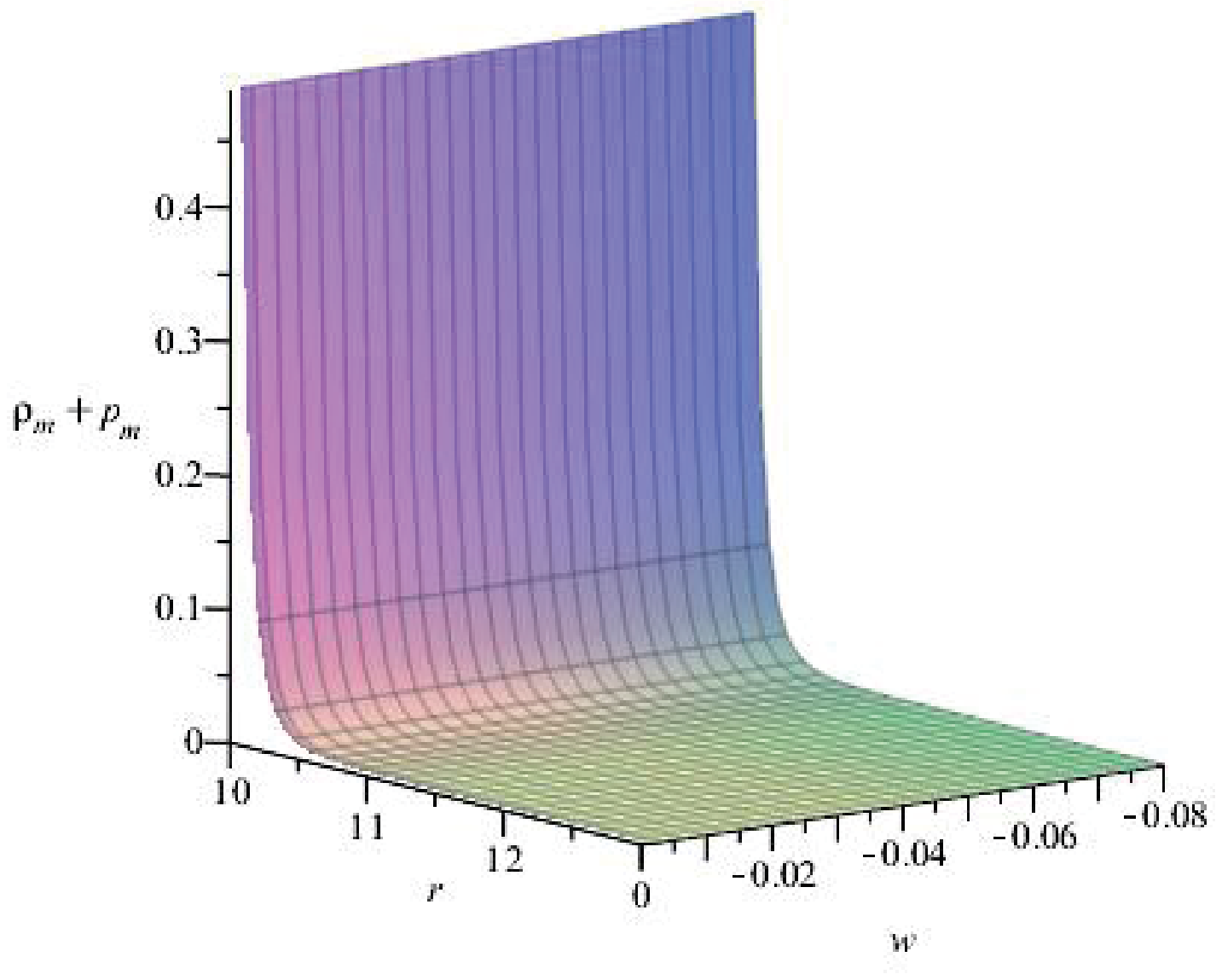,
width=0.5\linewidth}\\\epsfig{file=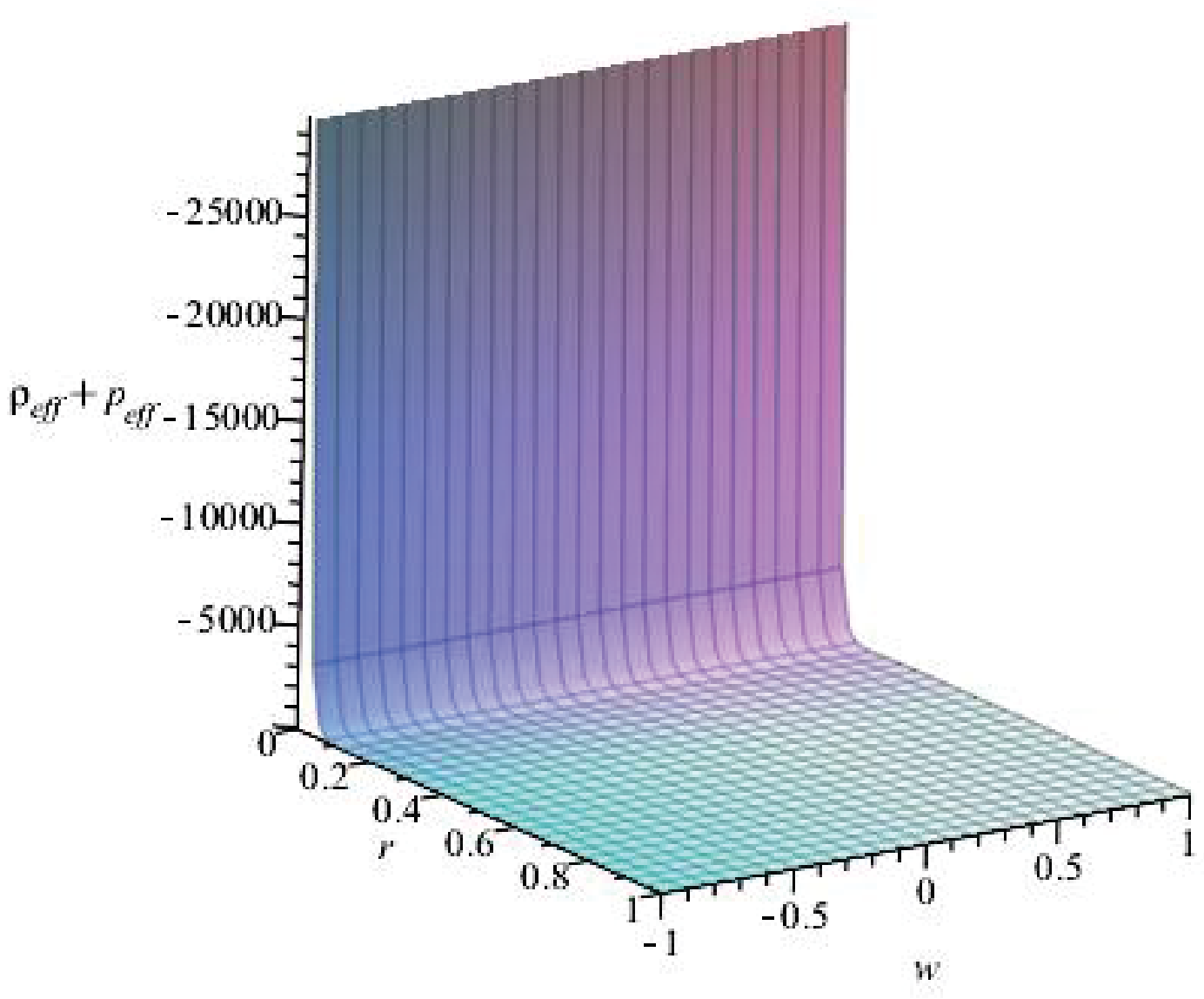,
width=0.5\linewidth}\caption{Plots of $\rho_m$, $\rho_m+p_m$ and
$\rho_{eff}+p_{eff}$ versus $r$.}
\end{figure}

Figure \textbf{4} shows that the model (\ref{A2}) follows the
stability condition for $0<\omega<-0.08$ whereas Figure \textbf{5}
represents the graphical behavior of the shape function. In upper
face, the left plot preserves the positivity of $h(r)$ while the
right plot ensures asymptotic flat geometry of WH. In lower face,
the left plot detects WH throat at $r_0=5.878$ whereas the right
plot indicates that $\frac{dh(r_0)}{dr}=0.1673<1$. For
Eqs.(\ref{10}) and (\ref{36}), we obtain
\begin{equation*}
\rho_{eff}+p_{eff}=\frac{k}{r^2(r-h(r))}+\frac{rh'(r)-h(r)}{r^3}.
\end{equation*}
To investigate the presence of realistic traversable WH, we
establish the graphical behavior of NEC and WEC corresponding to
perfect fluid as well as NEC relative to effective energy-momentum
tensor. Figure \textbf{6} indicates that $\rho_m+p_m\geq0$,
$\rho_m\geq0$ and $\rho_{eff}+p_{eff}<0$ for $1<\omega<-1$. Thus,
the physical existence of WH is assured in this case.

\subsection{Power-law $f(R)$ Model}

Here, we construct a WH solution with symmetry generator and
corresponding conserved quantity for $f(R)$ power-law model, i.e.,
$f(R)= f_0R^n,~n\neq0,1$. For this purpose, we solve
Eqs.(\ref{20})-(\ref{27}) leading to
\begin{equation*}
\alpha=Y_3(a,r),\quad\gamma=Y_1(r),\quad\delta=Y_2(r,R).
\end{equation*}
Inserting this solution into Eqs.(\ref{28})-(\ref{31}), we obtain
\begin{equation*}
Y_1(r)=0,\quad Y_3(a,r)=d_2,\quad
Y_2(r,R)=d_1R,\quad\beta=2(n-1)d_1+d_2-2\tau,_{_r},
\end{equation*}
where $d_1$ and $d_2$ represent arbitrary constants. For these
values, the coefficients of symmetry generator turn out to be
\begin{equation}
\alpha=d_2,\quad\beta=2(n-1)d_1+d_2-2\tau,_{_r},\quad\gamma=0,\quad\delta=d_1R.
\end{equation}
Substituting these coefficients in Eq.(\ref{32}) and assuming
$B=d_0$ and $\tau=\tau_0$, it follows that
\begin{eqnarray}\nonumber
b(r)&=&\int\frac{8d_3r^2+2a''r^2+4a'r'+a'^2r^2-4d_4}{r(4+a'r)}dr
\\\label{40}&-&\ln\left[-d_1+4\int\frac{e^{\int\frac{8r^2
+2a''r^2+4a'r'+a'^2r^2-4}{r(4+a'r)}dr}}{r(4+a'r)}dr\right].
\end{eqnarray}
The resulting coefficients of symmetry generator verifies the system
(\ref{20})-(\ref{31}) for $d_2=-2(n-1)d_1$. Under this condition,
the symmetry generator and associated first integral take the form
\begin{eqnarray*}
K&=&\tau_0\frac{\partial}{\partial
r}-2(n-1)d_1\frac{\partial}{\partial a}+d_1\frac{\partial}{\partial
R},\\\nonumber\Sigma&=&d_0-\tau_0\left[e^{\frac{a}{2}}e^{\frac{b}{2}}r^2(f-Rf_R
+\omega\rho_0a^{-\frac{(1+\omega)}{2\omega}}+2f_Rr^{-2})+\frac{e^{\frac{a}{2}}r^2}
{e^{\frac{b}{2}}}\right.\\\nonumber&\times&\left.\{f_R(2r^{-2}+2a'r^{-1})
+f_{RR}(a'R'+4R'r^{-1})\}\right]-2d_1(1-n)e^{\frac{a-b}{2}}(R'r^2\\\nonumber
&\times&f_{RR}+2rf_R)-d_1Rf_{RR}e^{\frac{a-b}{2}}(a'r^2+4r).
\end{eqnarray*}
Now, we solve the integral (\ref{40}) for constant and variable
forms of red-shift function and study WH geometry via shape
function.

\subsubsection*{Case I: $a(r)=k$}

For constant red-shift function, the integral (\ref{40}) reduces to
\begin{equation}\label{41}
b(r)=d_3r^2-d_4\ln r-\ln\left(\frac{-d_1r+e^{r^2}}{r}\right).
\end{equation}
This satisfies Eq.(\ref{32}) for
$\omega=1,\frac{1}{3},-\frac{1}{3},-1$ and
\begin{equation}\label{42}
\rho_0=-\frac{f_oe^{\frac{3\omega\ln d_1+4n\omega\ln2+\ln
d_1}{2\omega}}}{\omega d_1-(1+\omega)},\quad\omega\neq0.
\end{equation}
In this case, the shape function yields
\begin{equation}\label{43}
h(r)=r\left[1-d_4r\left(\frac{-d_1r+e^{r^2}}{r}\right)e^{-d_3r^2}\right].
\end{equation}
\begin{figure}
\centering{\epsfig{file=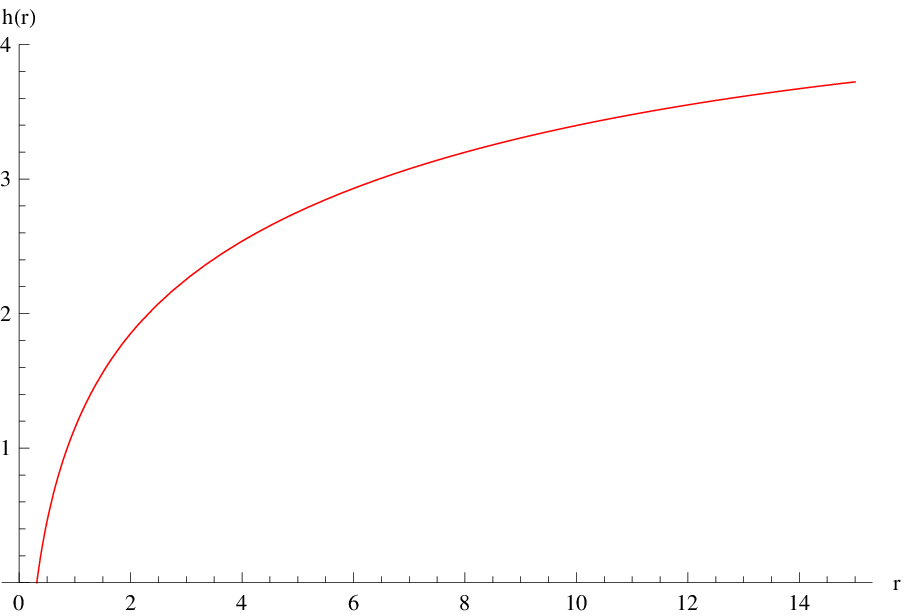,
width=0.35\linewidth}\epsfig{file=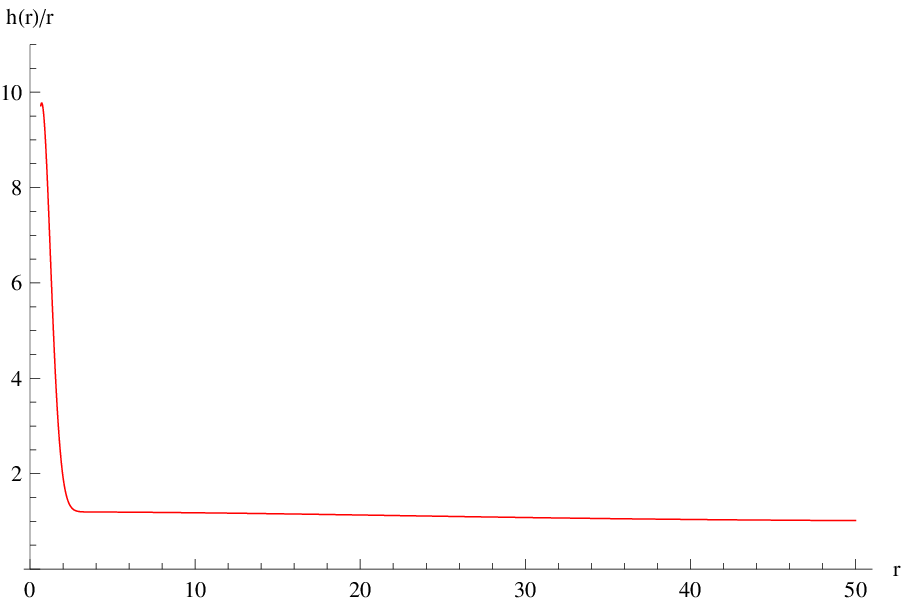,
width=0.35\linewidth}\\\epsfig{file=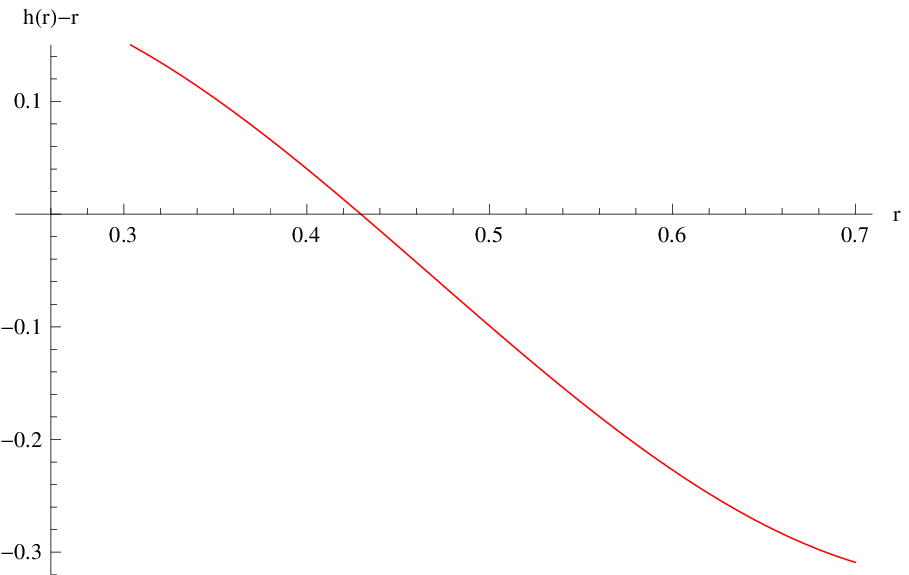,
width=0.35\linewidth}\epsfig{file=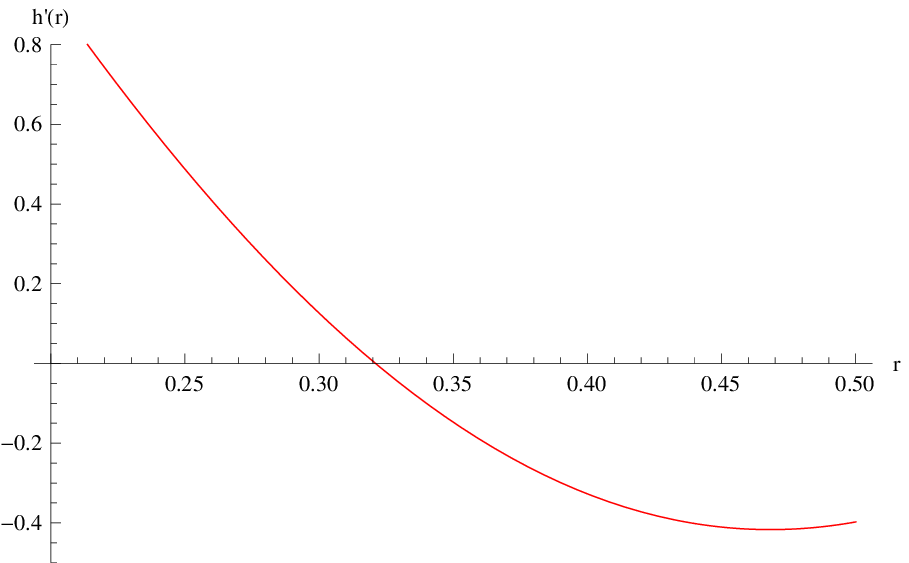,
width=0.35\linewidth}\caption{Plots of $h(r),~\frac{h(r)}
{r},~h(r)-r$ and $\frac{dh(r)}{dr}$ versus $r$ for $d_{_2}=16$,
$d_{_3}=1.001$, $d_{_4}=-0.2$ and $n=\frac{1}{2}$.}}
\end{figure}

We analyze WH geometry via shape function for $n=\frac{1}{2},2$ and
$n=4$. In upper face, the left and right plots of Figure \textbf{7}
show that $h(r)$ remains positive and asymptotic flat for
$n=\frac{1}{2}$. The lower left plot identifies WH throat at
$r_0=5.101$ and right plot satisfies the condition, i.e.,
$h'(r_0)=0.17<1$. In Figures \textbf{8} and \textbf{9}, the shape
function preserves its positivity condition and also admits
asymptotic flat geometry for both $n=2$ and $n=4$. The WH throat is
located at $r_0=0.23$ and $r_0=2.052$ for $n=2$ and $n=4$,
respectively. The derivative condition is also satisfied at throat,
i.e., $h'(r_0)=0.89<1$ and $h'(r_0)=-0.49<1$. The NEC relative to
effective energy-momentum tensor verifies $\rho_{eff}+p_{eff}<0$
while Figure \textbf{10} identifies $\rho_m\geq0$ and
$\rho_m+p_m\geq0$ for $n=0.5$. In case of $n=2$ and $n=4$, the
energy density and pressure corresponding to perfect fluid evolve in
the same way.
\begin{figure}\centering{\epsfig{file=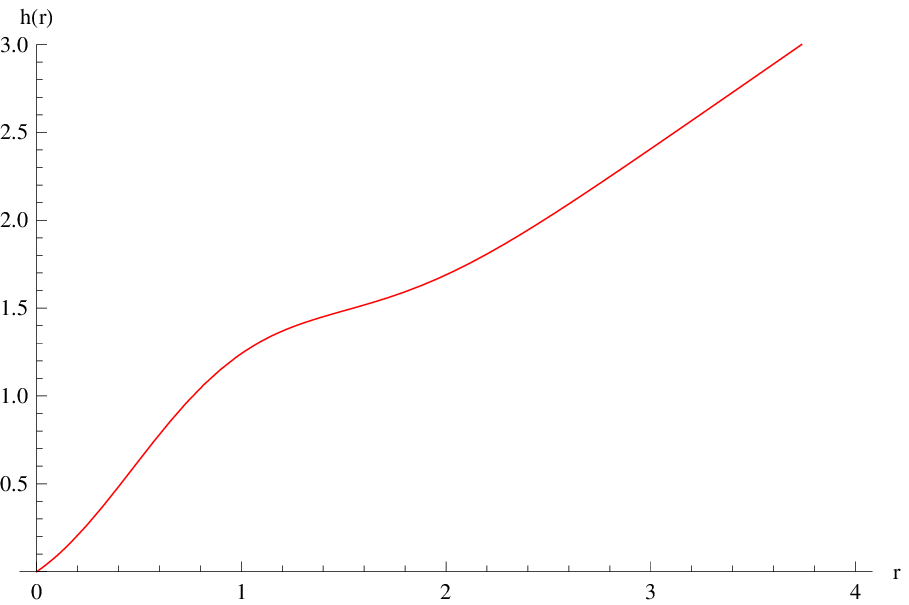,
width=0.35\linewidth}\epsfig{file=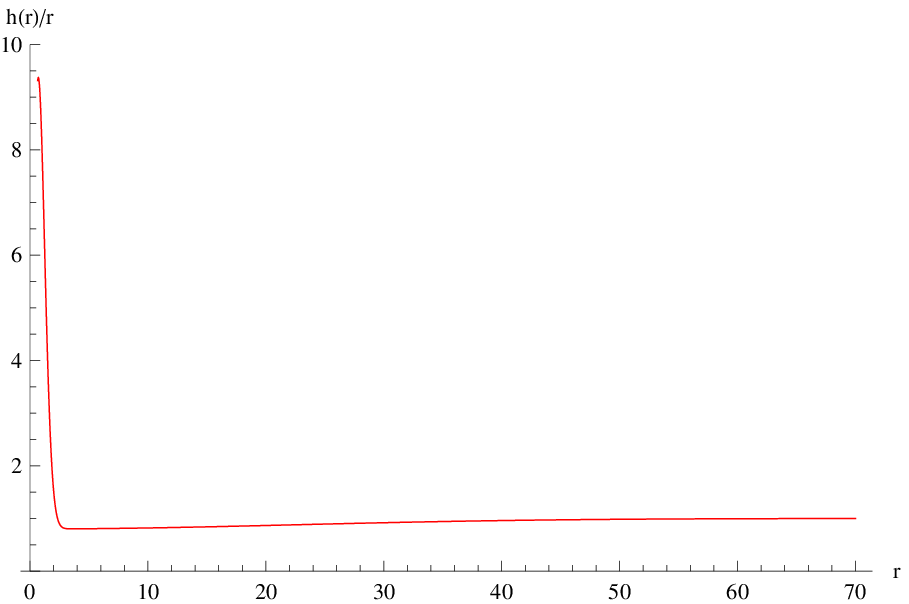,
width=0.35\linewidth}\\\epsfig{file=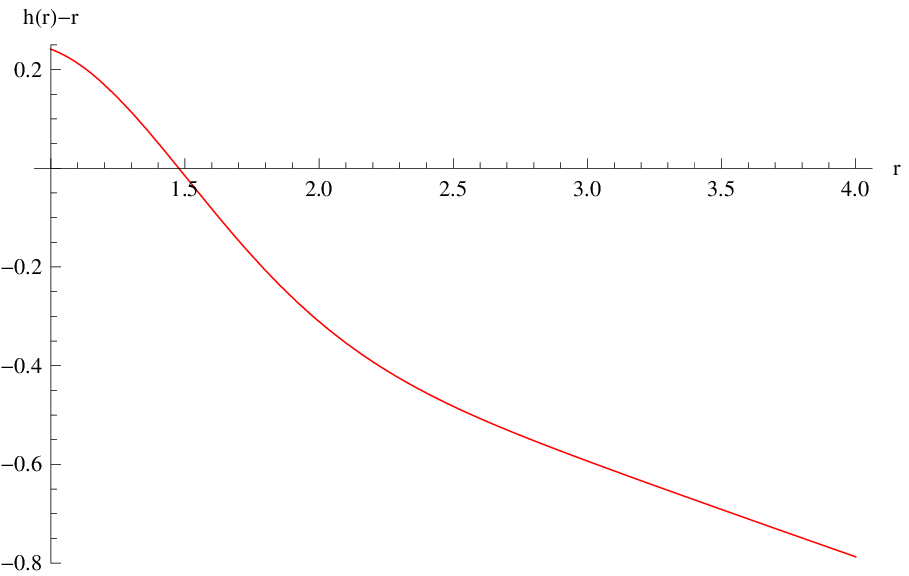,
width=0.35\linewidth}\epsfig{file=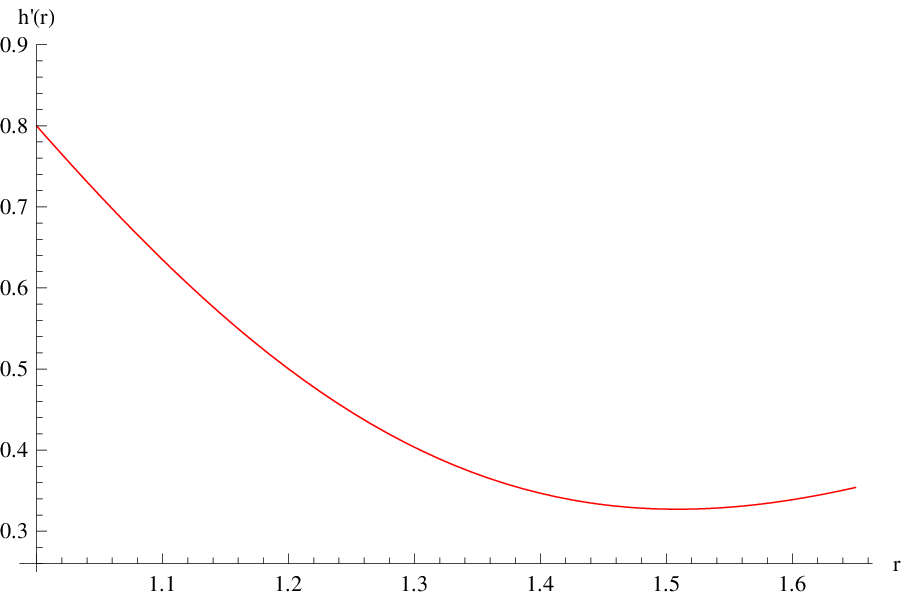,
width=0.35\linewidth}\caption{Plots of $h(r),~\frac{h(r)}
{r},~h(r)-r$ and $\frac{dh(r)}{dr}$ versus $r$ for $d_{_2}=-200$,
$d_{_3}=1.001$, $d_{_4}=0.2$ and $n=2$.}}
\end{figure}
\begin{figure}\centering{\epsfig{file=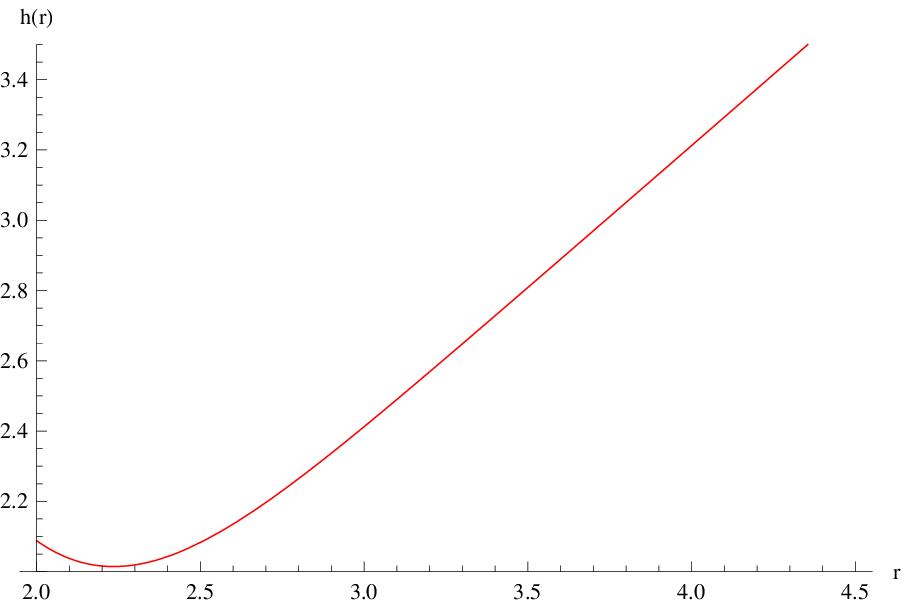,
width=0.35\linewidth}\epsfig{file=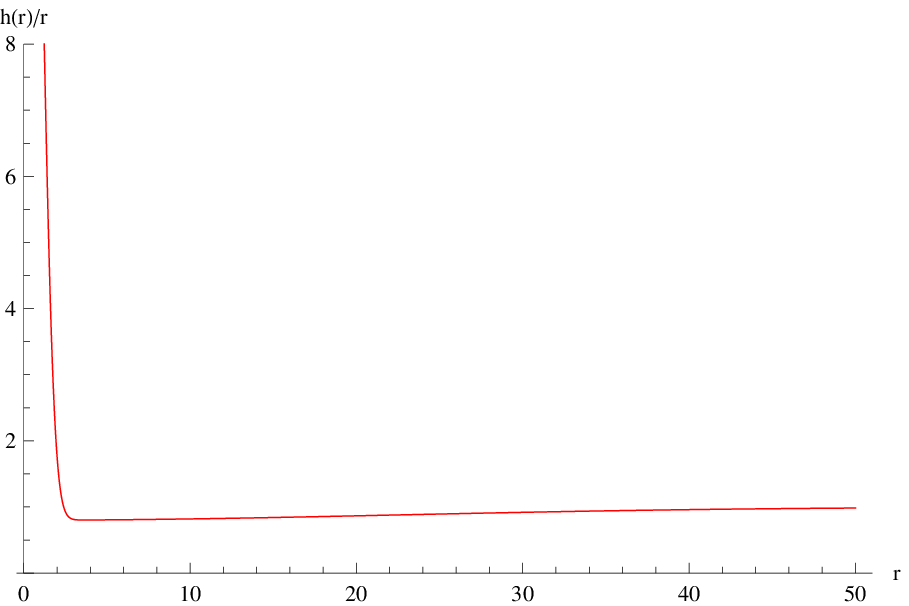,
width=0.35\linewidth}\\\epsfig{file=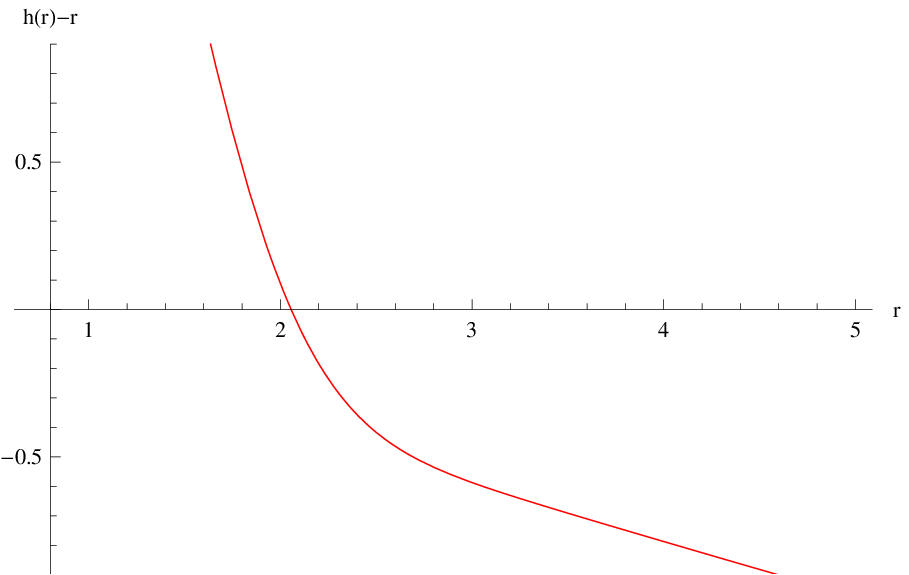,
width=0.35\linewidth}\epsfig{file=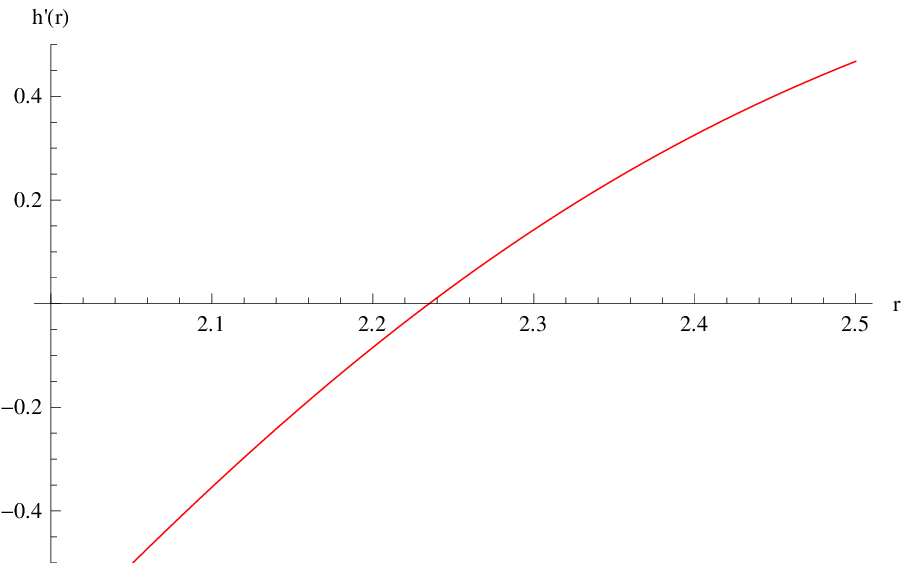,
width=0.35\linewidth}\caption{Plots of $h(r),~\frac{h(r)}
{r},~h(r)-r$ and $\frac{dh(r)}{dr}$ versus $r$ for $d_{_2}=-200$,
$d_{_3}=1.001$, $d_{_4}=0.2$ and $n=4$.}}
\end{figure}
\begin{figure}\centering{\epsfig{file=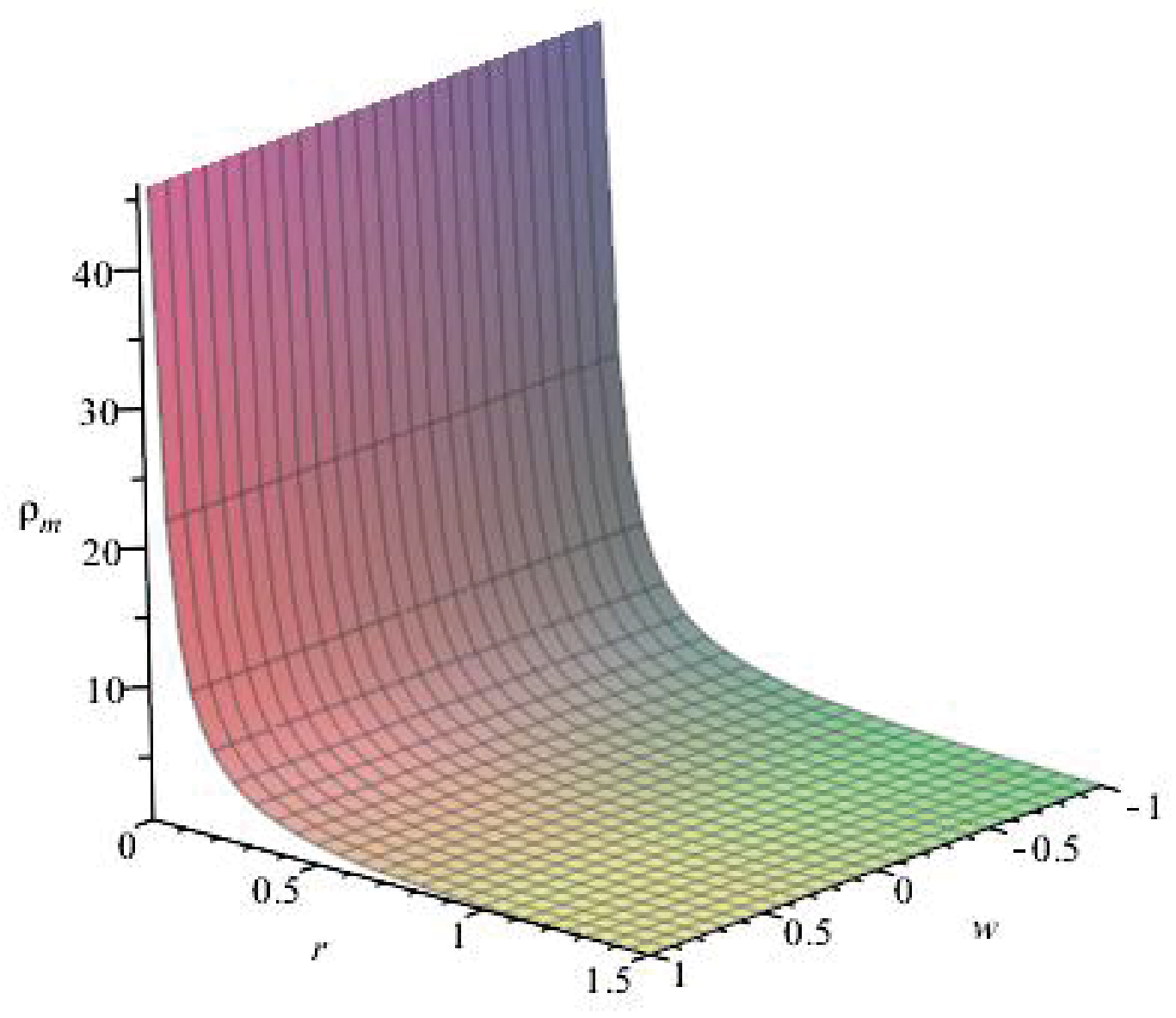,
width=0.65\linewidth}\epsfig{file=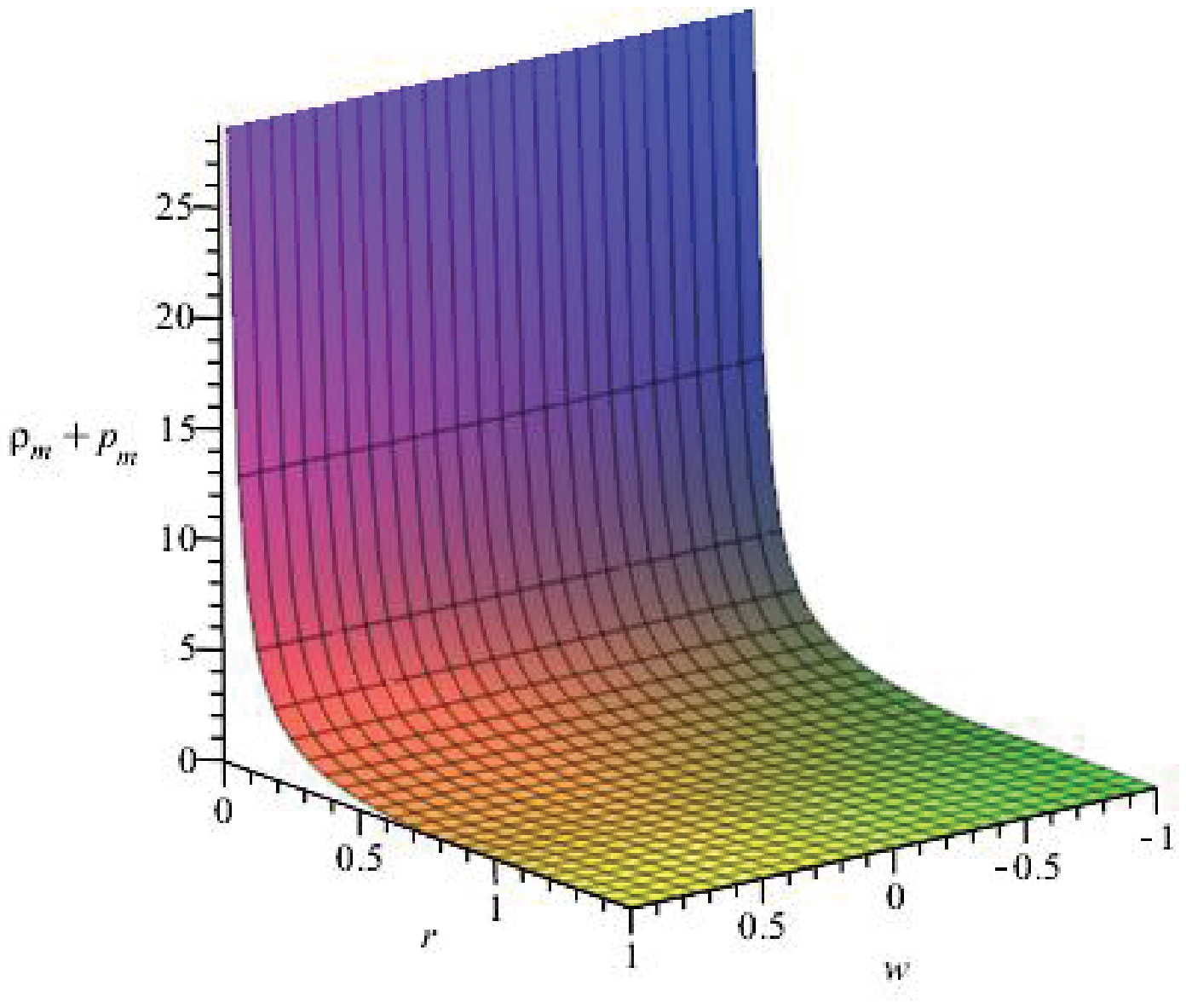,
width=0.45\linewidth}\caption{Plots of $\rho_m$ and $\rho_m+p_m$
versus $r$ for $n=0.5$.}}
\end{figure}

\subsubsection*{Case II: $a(r)=-k/r$}

Here we consider red-shift function to be $r$-dependent and solve
the integral (\ref{40}) implying that
\begin{eqnarray}\nonumber
b(r)&=&r^2-\frac{rd_1(1-n)}{2}+\frac{d_1^2(1-n)^2\ln(d_1(1-n)+4r)}{8}
+(d_1(1-n))^2\\\nonumber&\times&\left\{-\frac{1}{rd_1(1-n)}
+\frac{4\ln(d_1(1-n)+4r)}{(d_1(1-n))^2}-\frac{4\ln
r}{(d_1(1-n))^2}\right\}-\ln((1-n)\\\nonumber&\times&d_1+4r)
-\ln\left[4\int\frac{1}{4r+d_1(1-n)}\left(r^{-4}(d_1(1-n)
+4r)^{3+\frac{d_1^2(1-n)^2}{8}}\right.\right.\\\nonumber
&\times&\left.\left.e^{r^2-\frac{rd_1(1-n)}{2}+\frac{d_1(1-n)}{r}}
\right)dr-d_1\right].
\end{eqnarray}
This solution satisfies Eq.(\ref{32}) for $\omega=-1$. The shape
function of WH takes the form
\begin{eqnarray*}
&&\frac{h(r)}{r}=\left(1-r^{4}(d_1(1-n)+4r)^{-3-\frac{d_1^2(1-n)^2}{8}}
e^{-r^2+\frac{rd_1(1-n)}{2}-\frac{d_1(1-n)}{r}}\left[\int\{4r+d_1\right.
\right.\\\nonumber&&\times\left.\left.(1-n)\}^{-1}\left(r^{-4}(d_1(1-n)
+4r)^{3+\frac{d_1^2(1-n)^2}{8}}e^{r^2-\frac{rd_1(1-n)}{2}+\frac{d_1(1-n)}{r}}
\right)dr-d_1\right]\right).
\end{eqnarray*}
When red-shift function is not constant ($a'(r)\neq0$), then the
geometry of WH cannot be analyzed for $f(R)$ power-law model due to
the complicated forms of $b(r)$ and $h(r)$.

\subsection{Exponential Model}

In this section, we consider another example of viable $f(R)$ model,
i.e., exponential model to realize the existence of realistic
traversable WH. The simplest version of this model is proposed as
\cite{exp1}
\begin{equation}\label{exp1}
f(R)=R-2\Lambda(1-e^{-\frac{R}{R_0}}),
\end{equation}
where $\Lambda$ denotes cosmological constant while $R_0$ defines
curvature parameter. If $R\gg R_0$, then the corresponding model
recovers standard cosmological constant cold dark matter model. To
formulate WH solution, we first solve the system of
Eqs.(\ref{20})-(\ref{32}) for the model (\ref{exp1}) which leads to
the following coefficients of symmetry generator and boundary term
\begin{eqnarray*}
\alpha&=&0,\quad\beta=\frac{4\Lambda\chi_1}{R_0},\quad\gamma=0,\quad
\delta=\chi_1(R_0e^{\frac{R}{R_0}}-2\Lambda),\quad\tau=\tau_0,\\\nonumber
B&=&\frac{2e^{\frac{a+b}{2}}\Lambda\chi_1}{R_0^2}\left[-\frac{2r^3R_0\Lambda}{3}\left(1-
e^{-\frac{R}{R_0}}-\frac{2R}{R_0}\right)+\frac{r^3R_0(1-RR_0)}{3}+4r\right.
\\\nonumber&\times&\left.(R_0-2\Lambda e^{-\frac{R}{R_0}})\right]+\chi_2,
\end{eqnarray*}
where $\chi_1$ and $\chi_2$ represent arbitrary constants. These
solutions satisfy the system for $\omega=\rho_0=-1$ and the
following constraint
\begin{equation}\label{exp2}
e^{\frac{R}{R_0}}r^2R_0^2-2r^2R_0\Lambda+4r^2R\Lambda-24\Lambda=0.
\end{equation}
Now we determine the coefficient of radial component of the metric
(\ref{6}) using this constraint with Eq.(\ref{12}) for both constant
as well as variable forms of red-shift function and study WH
geometry via shape function.

\subsubsection*{Case I: $a(r)=k$}
\begin{figure}\centering{\epsfig{file=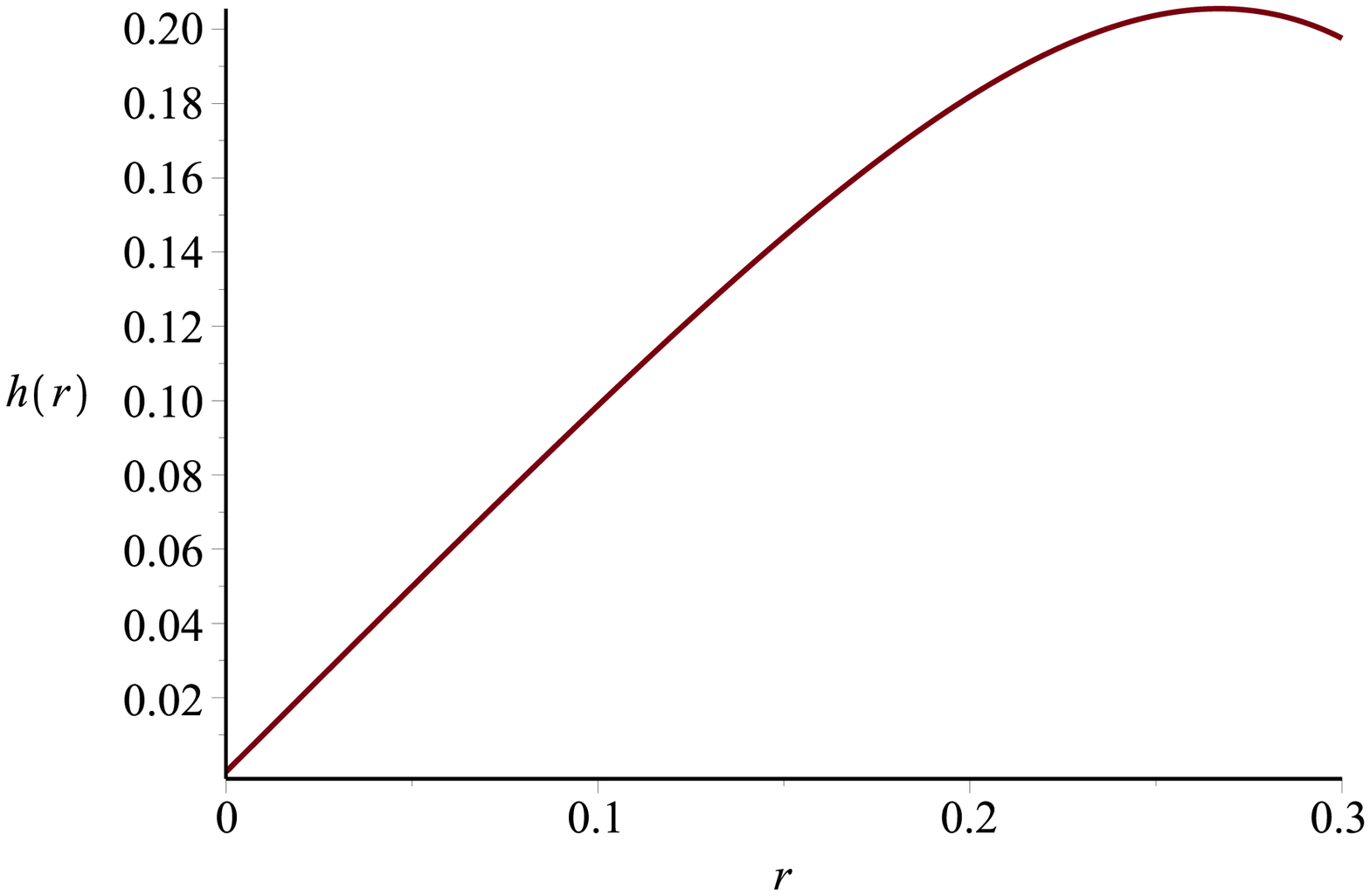,
width=0.45\linewidth}\epsfig{file=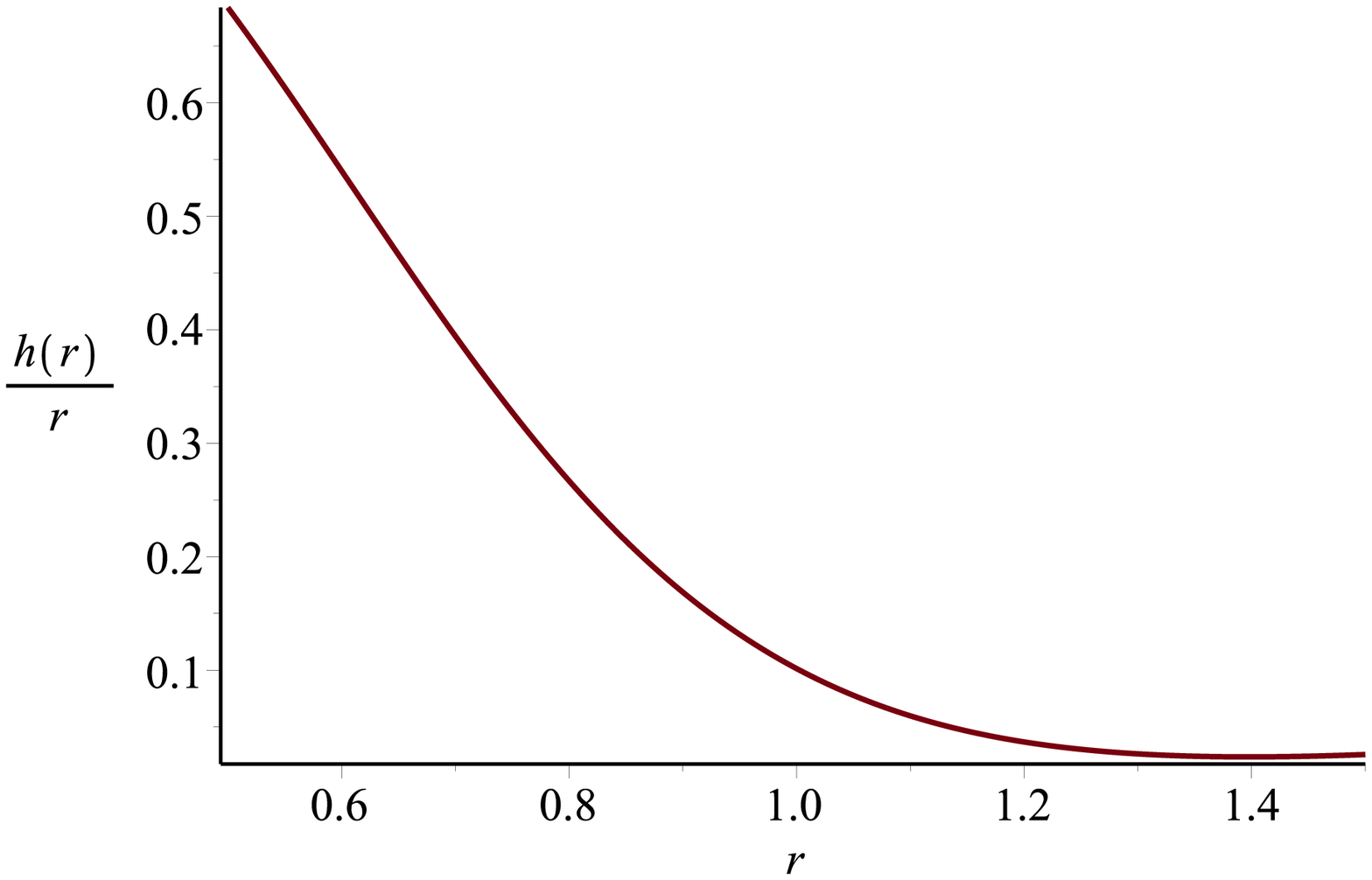,
width=0.45\linewidth}\\\epsfig{file=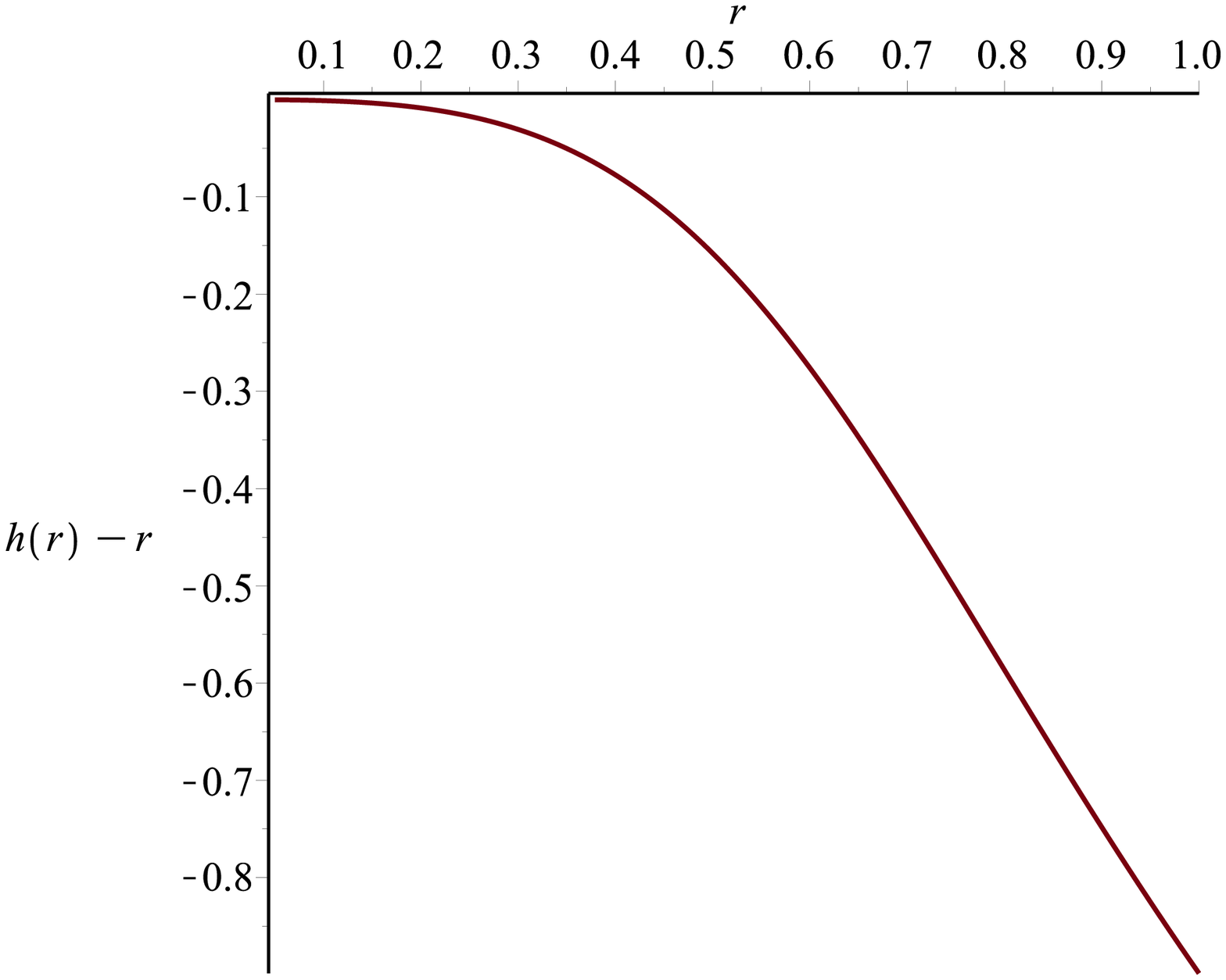,
width=0.45\linewidth}\epsfig{file=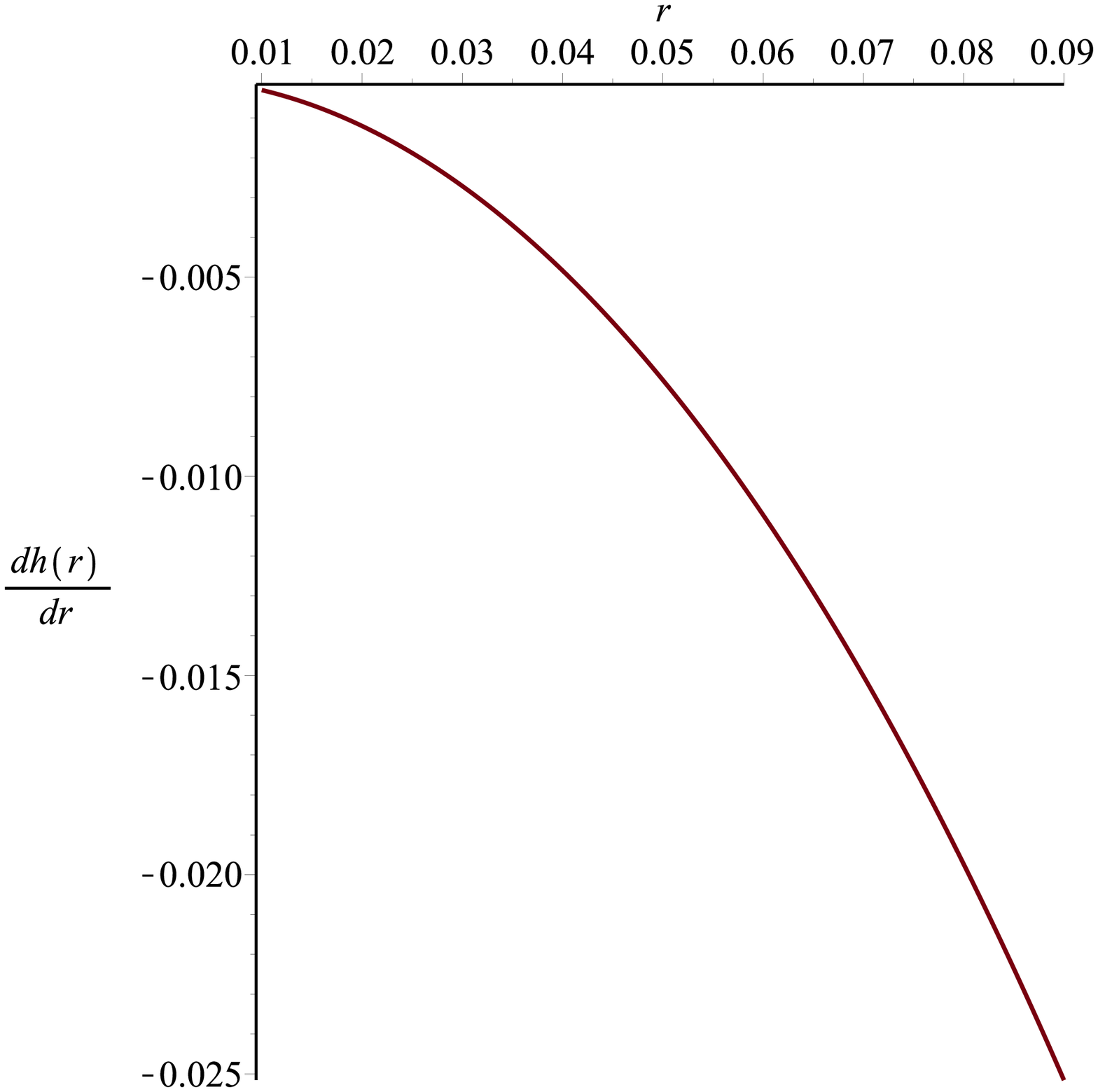,
width=0.45\linewidth}\caption{Plots of $h(r),~\frac{h(r)}
{r},~h(r)-r$ and $\frac{dh(r)}{dr}$ versus $r$ for $\chi_{_4}=-200$,
$R_0=-0.95=\Lambda$ and $k=0.005$.}}
\end{figure}
\begin{figure}\centering{\epsfig{file=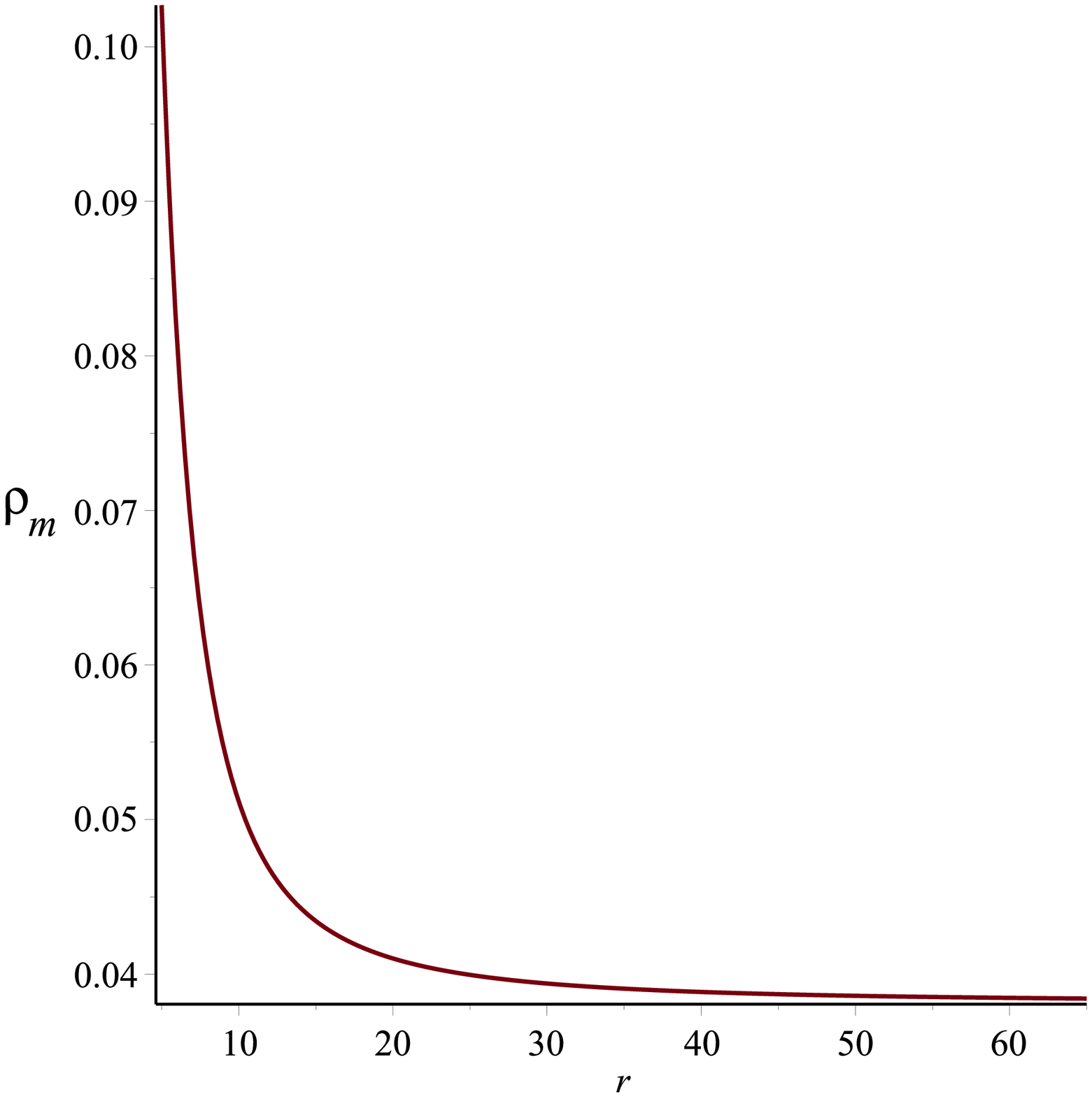,
width=0.45\linewidth}\epsfig{file=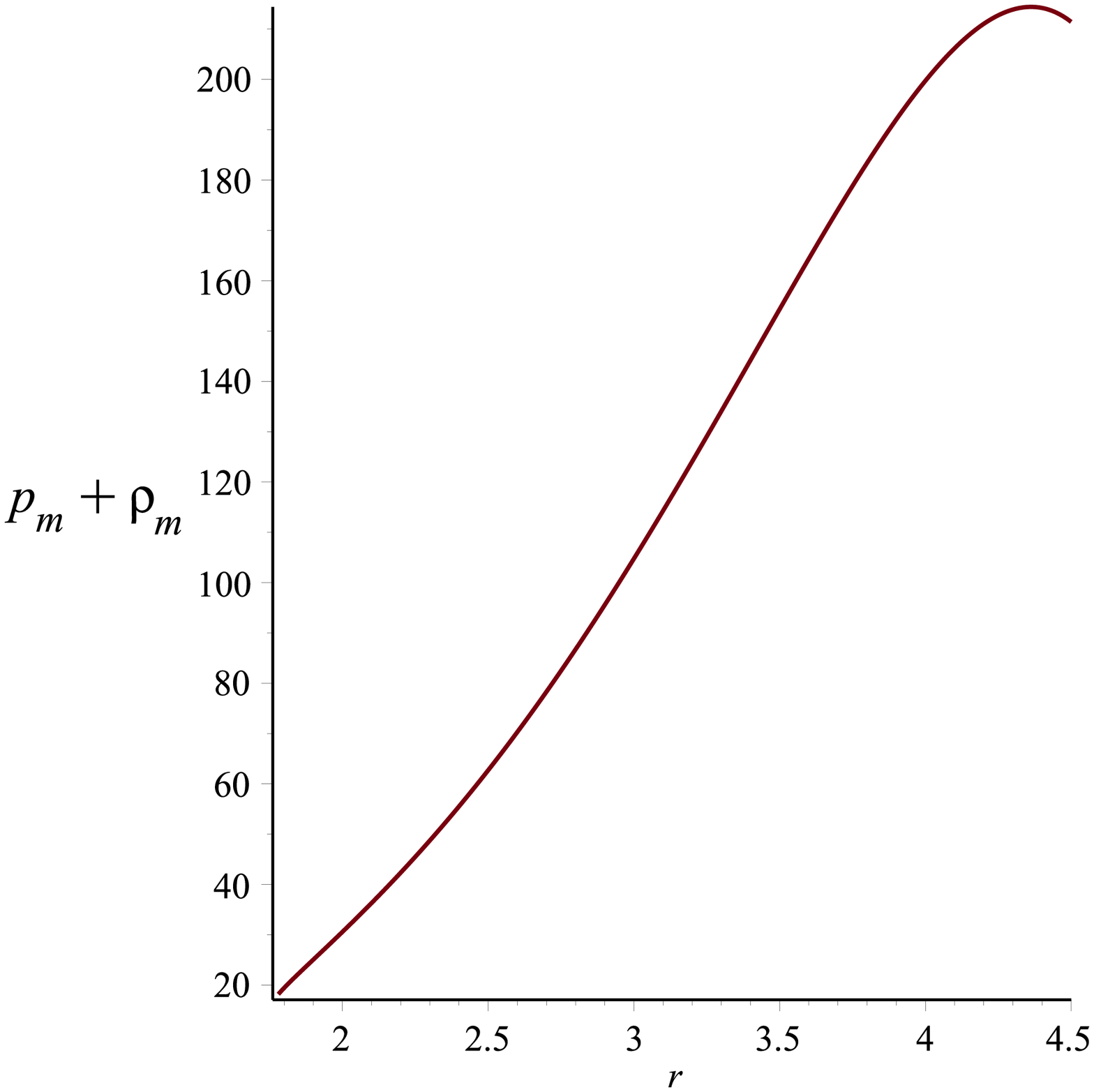,
width=0.45\linewidth}\caption{Plots of $\rho_m$ and $\rho_m+p_m$
versus $r$.}}
\end{figure}

In this case, we obtain
\begin{eqnarray}\nonumber
e^{b(r)}&=&-(4(-2R_0r^2+(R_0r^2+12r^4)\exp((1/2)(12+R_0r^2)/(R_0r^2))\\\nonumber
&-&48r^4\chi_4+24(1-\chi_4)))\{(r^2((5r^4R_0^2-2r^4R_0
-4R_0r^2)\\\nonumber&\times&\exp((1/2)(12+R_0r^2)(R_0r^2)^{-1})
-6r^4R_0^2+48r^2\chi_4-48r^2\\\label{41e}&
-&120R_0r^2\chi_4+104R_0r^2-96+96\chi_4))\}^{-1}.
\end{eqnarray}
From this expression, we formulate shape function through
$h(r)=r[1-e^{-b(r)}]$ and analyze the WH geometry graphically. In
Figure \textbf{11}, the upper face indicates that the shape function
is positively increasing while the corresponding geometry is found
to be asymptotically flat as $h(r)/r\rightarrow0$ when
$r\rightarrow\infty$. In the lower face, the left plot indicates
that the WH throat exists at $r_0=0.05$ and also preserves the
condition, i.e., $h(0.05)=0.05$ while the right plot shows that
$h'(r_0)=-0.007<1$. Since the red-shift function is constant
therefore, the traversable nature of the constructed WH solution is
preserved by the violation of effective NEC, i.e.,
$p_{eff}+\rho_{eff}<0$. Figure \textbf{12} evaluates the criteria
for physically viable WH as $\rho_m>0$ and $p_m+\rho_m>0$.

\subsubsection*{Case II: $a(r)=-k/r$}

Using Eqs.(\ref{12}) and (\ref{exp2}), it follows that
\begin{eqnarray}\nonumber
e^{b(r)}&=&-(4(24+48kr^2-2R_0r^2-4kr^4R_0-12(r+4)kr^2\chi_4\\\nonumber&-&
24\chi_4(1+2r^4))+(2kr^4R_0+3r^3k+R_0r^2+12r^4)\\\nonumber&\times&
\exp((1/2)(12+R_0r^2)/(R_0r^2)))\{r^2
((-2r^4R_0+5r^4R_0^2-4R_0r^2)\\\nonumber&\times&\exp((1/2)
(12+R_0r^2)/(R_0r^2))-(6r^2R_0+104)R_0r^2-48(r^2+2)
\\\nonumber&-&(120R_0r^2-48r^2+96)\chi_4\}^{-1}.
\end{eqnarray}
Inserting the above expression in $h(r)=r[1-e^{-b(r)}]$, we
construct WH solution relative to variable but finite red-shift
function whose graphical interpretation is given in Figure
\textbf{13}. Both plots of the upper and lower panels indicate that
the constructed WH follows asymptotic flat geometry whose throat is
located at $r_0=0.01$ and $h'(0.01)=-0.001<1$. In order to analyze
the presence of repulsive gravitational effects at throat, we study
the behavior of effective NEC in Figure \textbf{14} which ensures
that the sum of $p_{eff}$ and $\rho_{eff}$ remains negative. Thus,
the constructed WH is found to be traversable. Both plots of Figure
\textbf{15} shows that the WH is physically viable as NEC and WEC
corresponding to ordinary matter are preserved.
\begin{figure}\centering{\epsfig{file=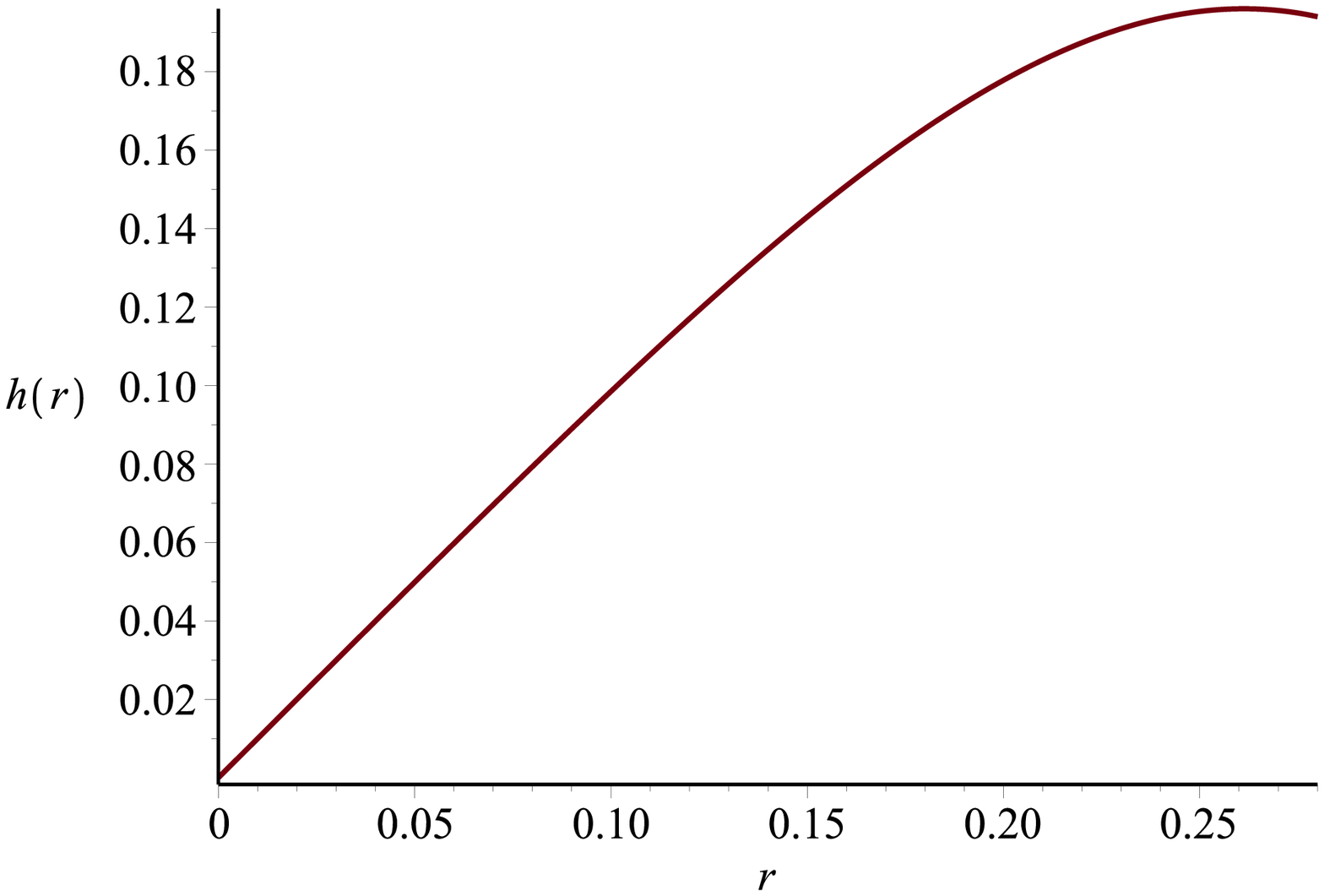,
width=0.45\linewidth}\epsfig{file=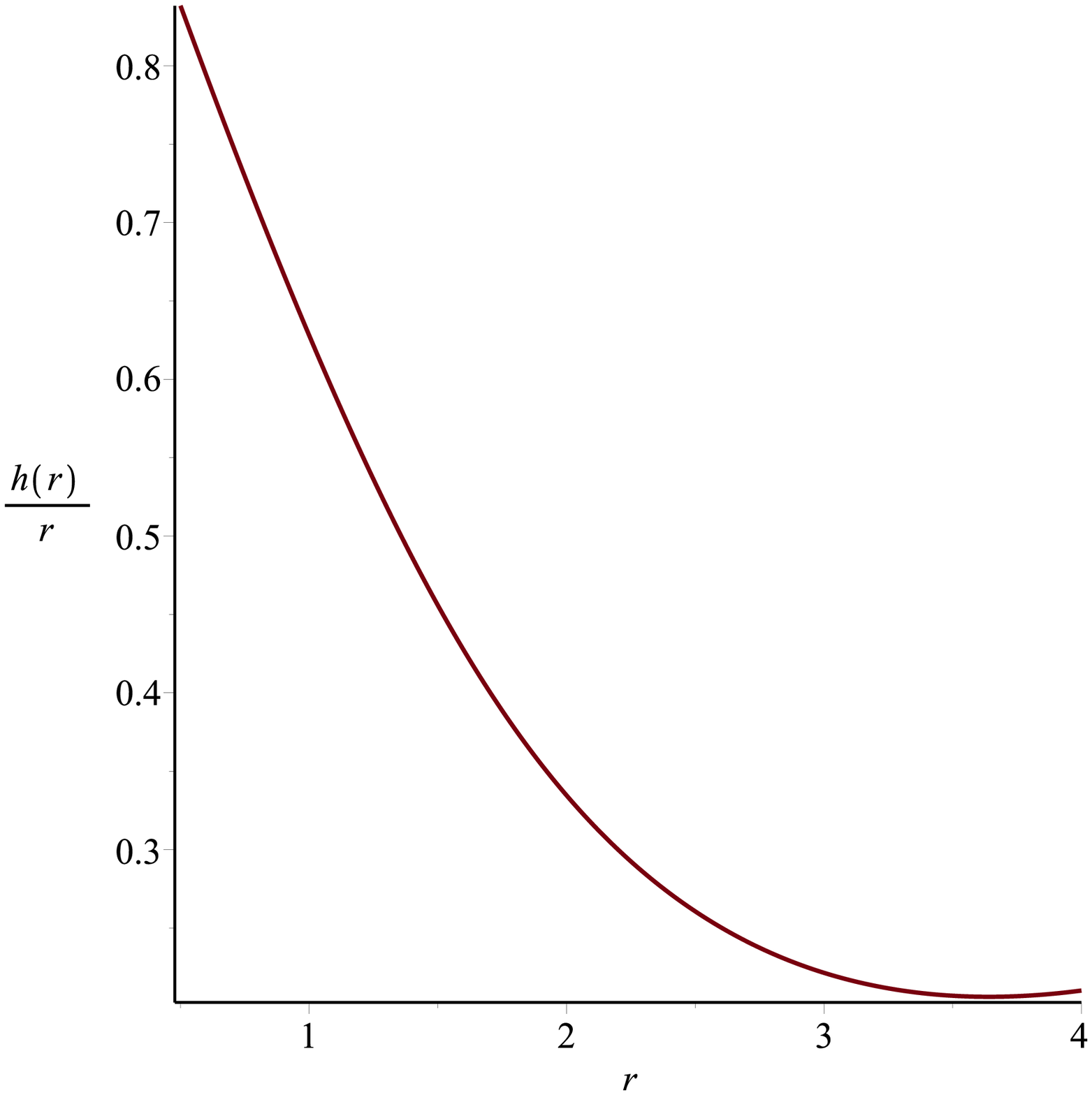,
width=0.4\linewidth}\\\epsfig{file=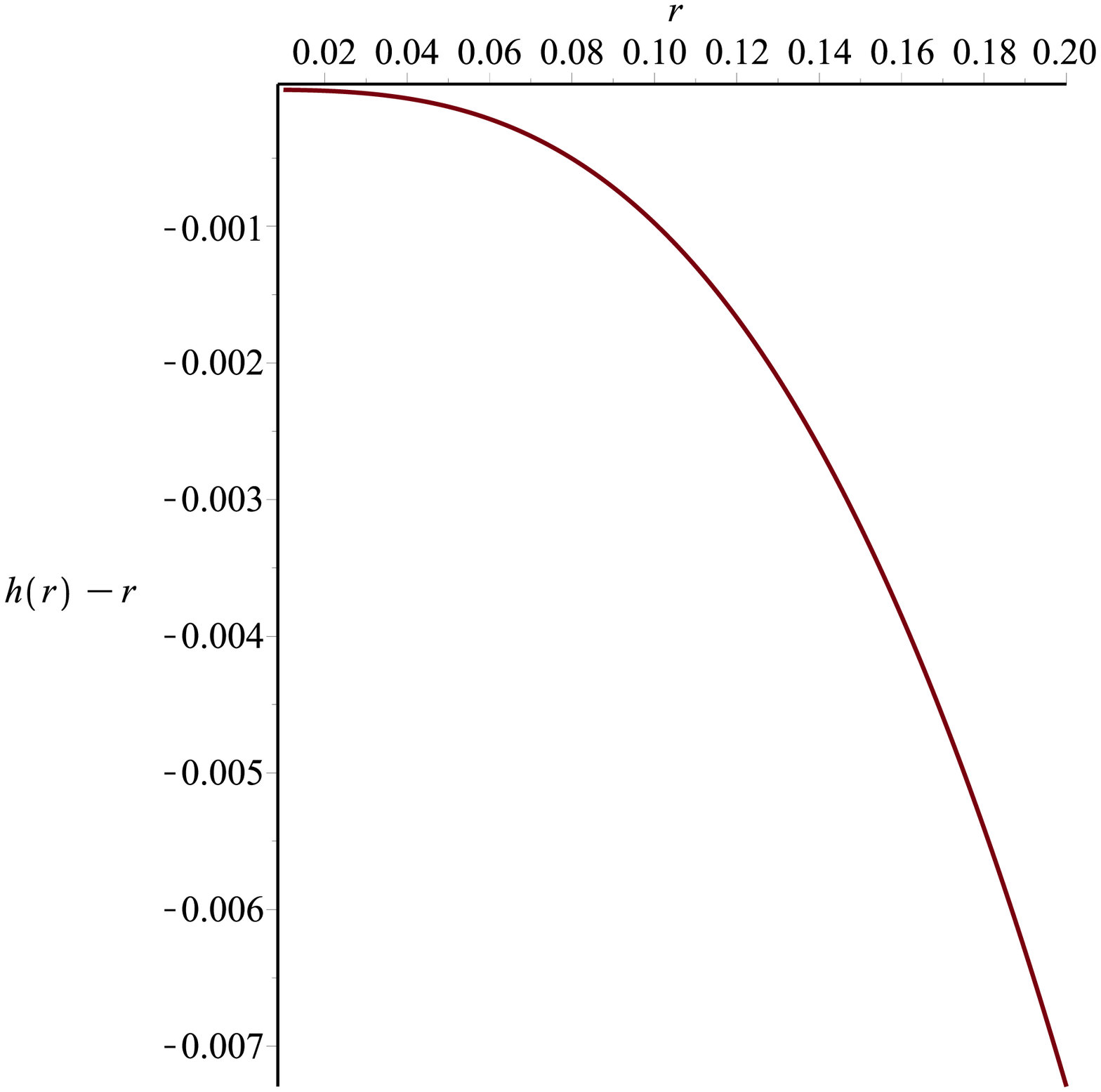,
width=0.45\linewidth}\epsfig{file=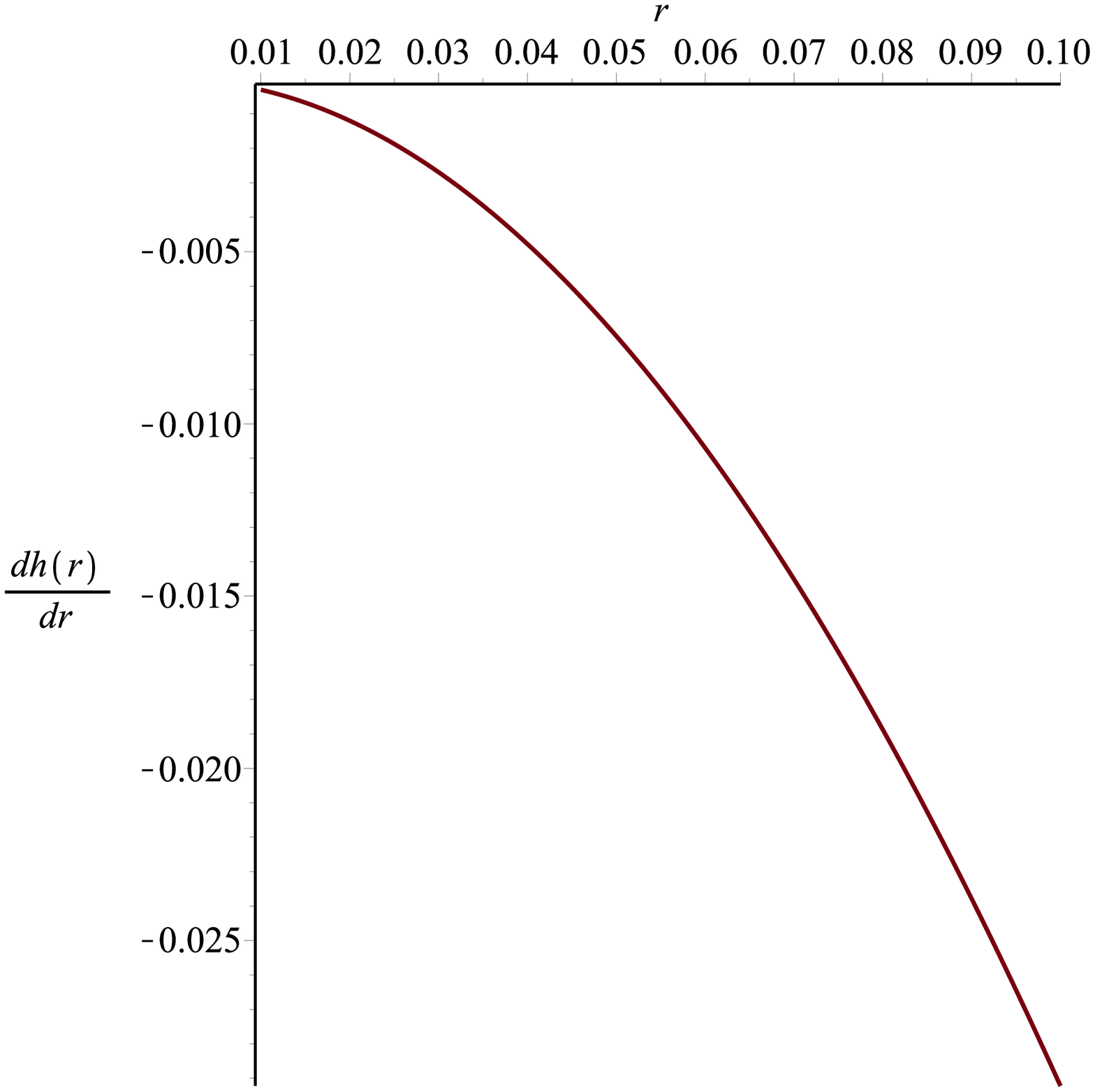,
width=0.45\linewidth}\caption{Plots of $h(r),~\frac{h(r)}
{r},~h(r)-r$ and $\frac{dh(r)}{dr}$ versus $r$ for
$\chi_{_4}=-0.20$, $R_0=-0.95=\Lambda$ and $k=2$.}}
\end{figure}
\begin{figure}\centering{\epsfig{file=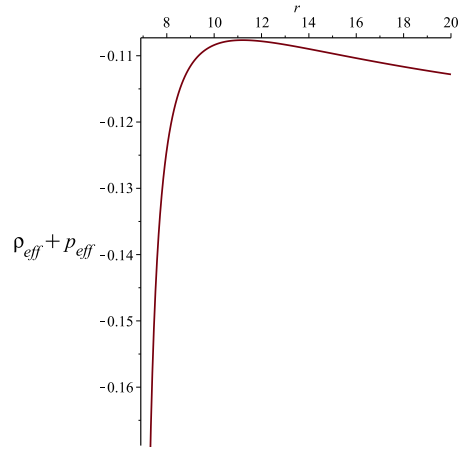,
width=0.45\linewidth}\caption{Evolution of $\rho_{eff}+p_{eff}$
versus $r$.}}
\end{figure}\begin{figure}\centering{\epsfig{file=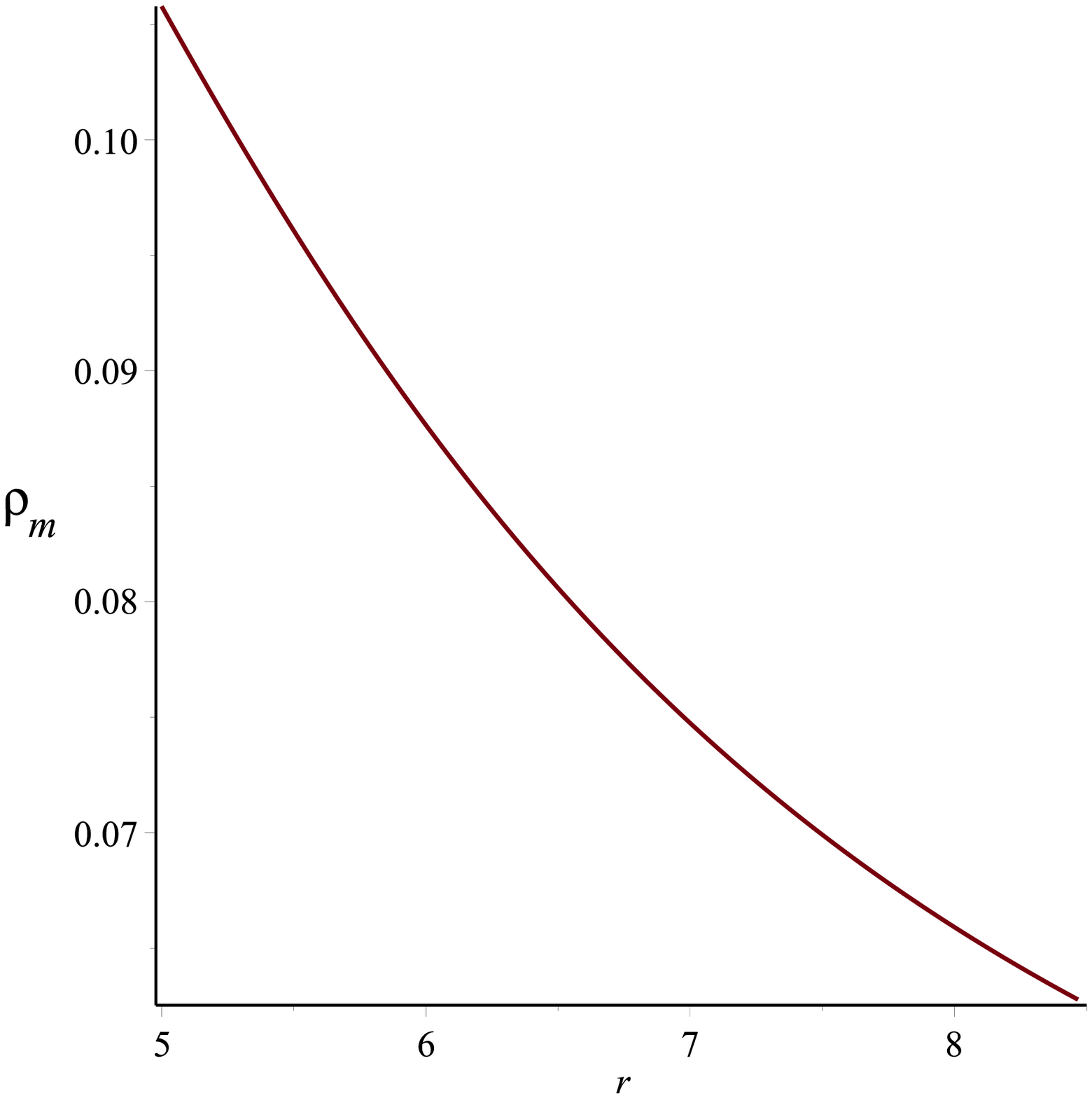,
width=0.45\linewidth}\epsfig{file=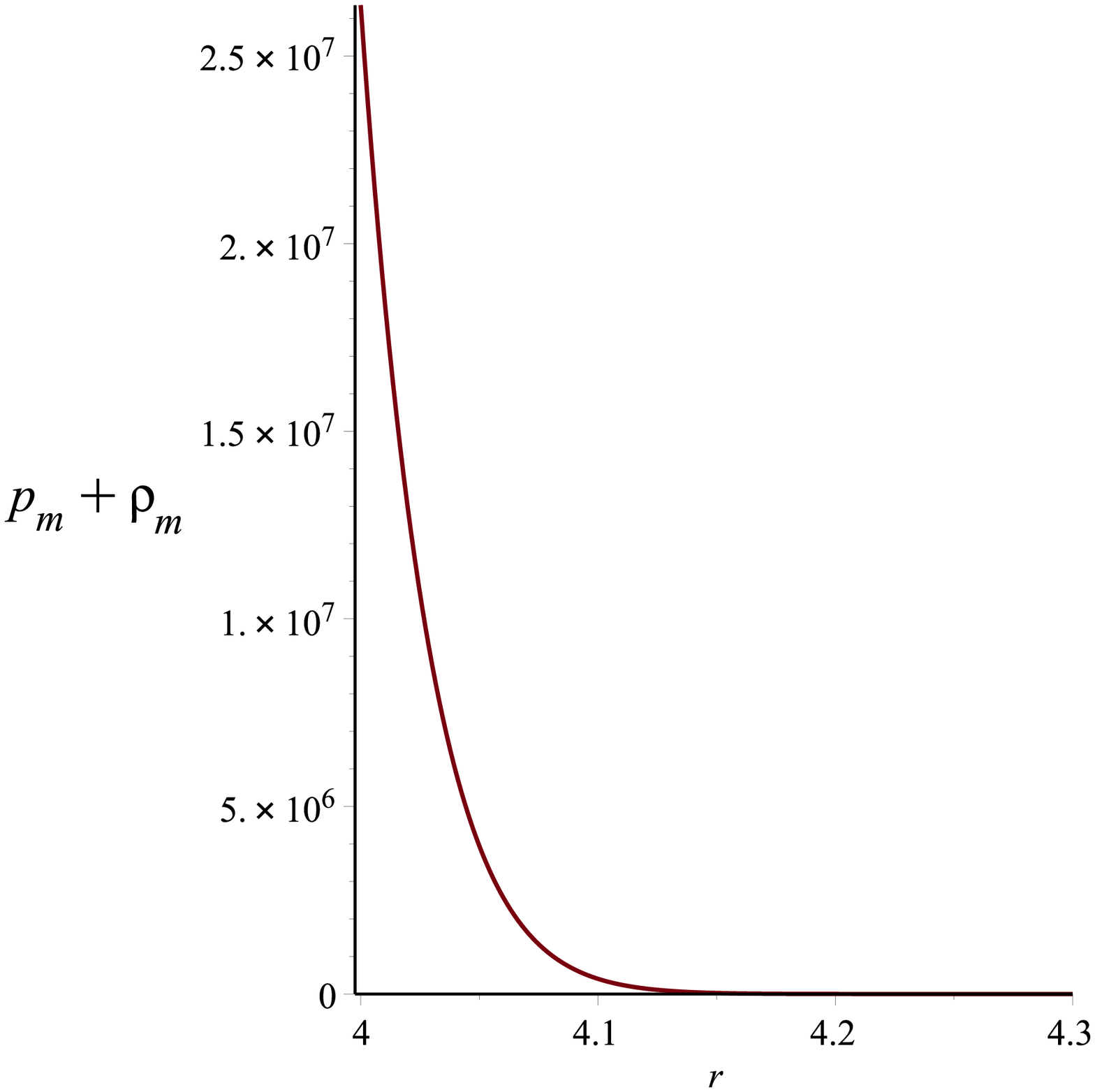,
width=0.45\linewidth}\caption{Plots of $\rho_m$ and $\rho_m+p_m$
versus $r$.}}
\end{figure}

\section{Stability Analysis}

Here we discuss the stability of WH solutions relative to both
constant as well as variable red-shift function via
Tolman-Oppenheimer-Volkov (TOV) equation. For isotropic fluid
distribution, the radial component of Bianchi identity
($\nabla_{\mu}T^{\mu\nu}=0$) defines TOV equation as
\begin{equation}\label{N1}
\frac{dp_m}{dr}+\frac{a'(r)}{2}\left(p_m+\rho_m\right)=0.
\end{equation}
The conservation of energy-momentum tensor relative to high order
curvature terms leads to
\begin{eqnarray}\label{N2}
T^{'(c)}_{11}+\frac{a'}{2}\left(T^{(c)}_{00}+T^{(c)}_{11}\right)
-\frac{M'}{M}\left(f_R''-\frac{f_R'}{e^{b(r)}}\left\{\frac{b'}{2}
+\frac{M'}{2M}\right\}\right)=0.
\end{eqnarray}
Combining Eq.(\ref{N1}) and (\ref{N2}), it follows that
\begin{eqnarray}\label{N3}
p'_{(eff)}+\frac{a'(r)}{2}\left(p_{eff}+\rho_{eff}\right)
-\frac{M'}{M}\left(f_R''-\frac{f_R'}{e^{b(r)}}\left\{\frac{b'}{2}
+\frac{M'}{2M}\right\}\right)=0,
\end{eqnarray}
where $p_{eff}=p_m+T^{(c)}_{11}$ and
$\rho_{eff}=\rho_m+T^{(c)}_{00}$. This equation determines the fate
of the WH as it can be expressed as a combination of hydrostatic
$\mathcal{F}_h$ and gravitational force $\mathcal{F}_g$. Using
Eq.(\ref{N3}), these forces take the following form
\begin{eqnarray*}
\mathcal{F}_h&=&p'_{(eff)}=\frac{d}{dr}(p_m+T^{(c)}_{11}),\\\nonumber
\quad\mathcal{F}_g&=&\frac{\mathcal{M}_{eff}e^{\frac{a-b}{2}}}{r^2}
\left(p_{eff}+\rho_{eff}\right)-\frac{M'}{M}\left(f_R''-\frac{f_R'}
{e^{b(r)}}\left\{\frac{b'}{2}+\frac{M'}{2M}\right\}\right),
\end{eqnarray*}
where $\mathcal{M}_{eff}=\frac{a'r^2e^{\frac{b-a}{2}}}{2}$ denotes
effective gravitational mass. The null effect
($\mathcal{F}_h+\mathcal{F}_g=0$) of these dynamical forces leads to
stable state of a WH.
\begin{figure}\centering{\epsfig{file=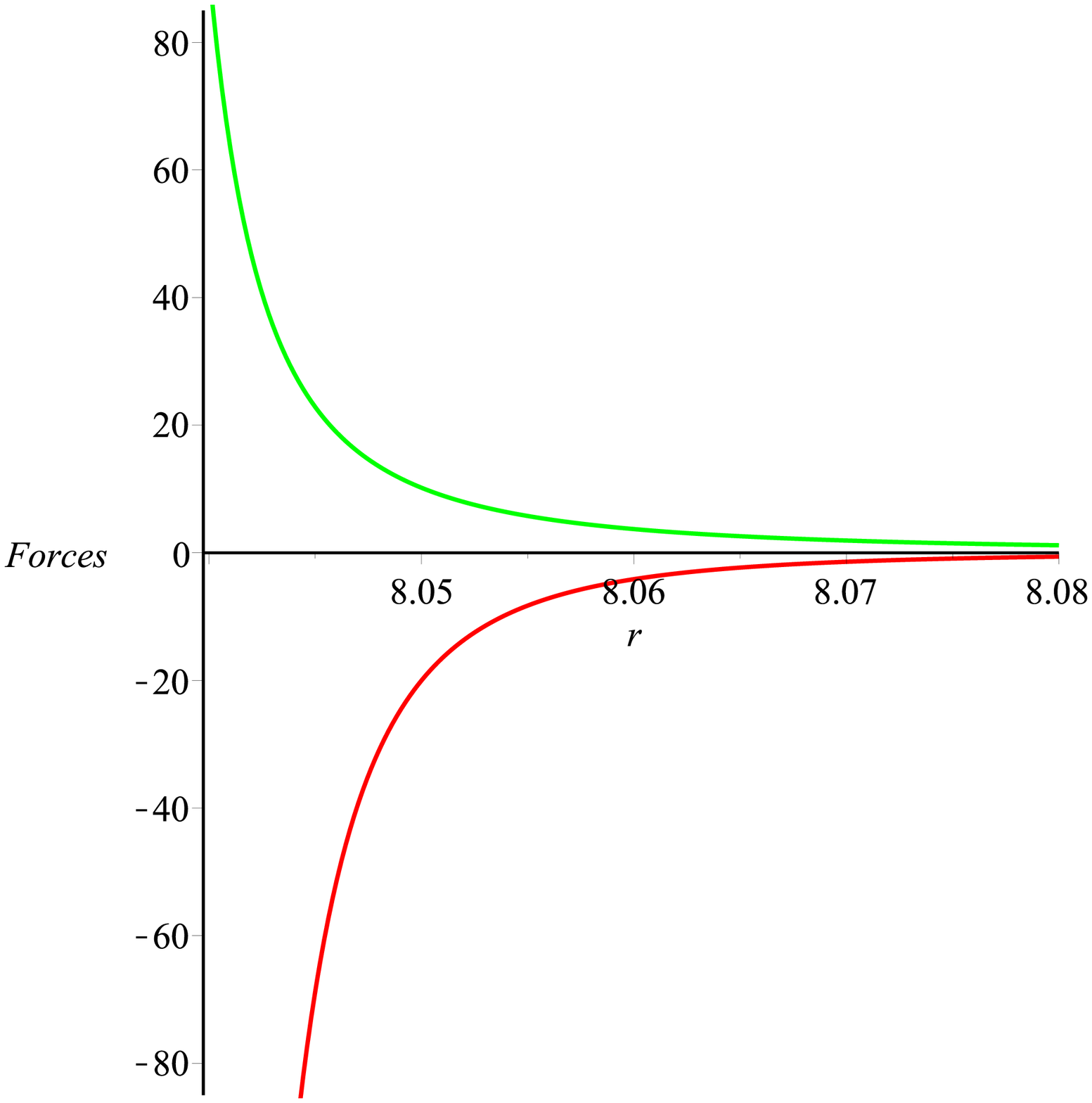,
width=0.45\linewidth}\epsfig{file=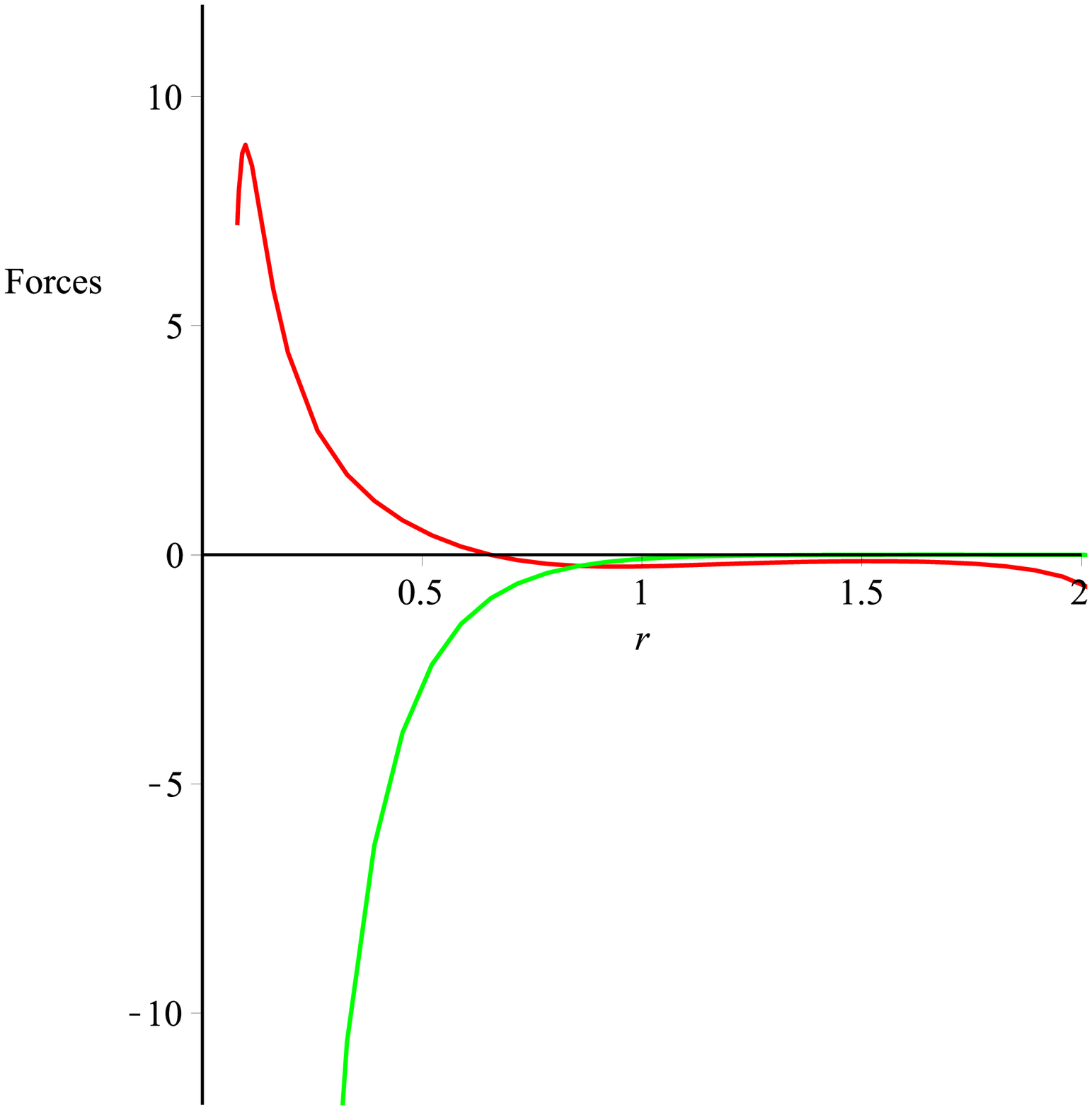,
width=0.45\linewidth}\caption{Plots of $\mathcal{F}_g$ (green) and
$\mathcal{F}_h$ (red) versus $r$ for $a(r)=k$ (left) and $a(r)=-k/r$
(right) for $c_{_2}=5$, $c_{_4}=0.01$, $c_{_5}=-0.35$, $c_{_6}=0.1$,
$c_{_7}=-0.25$, $\rho_0=-0.01$ and $k=0.5$.}}
\end{figure}
\begin{figure}\centering\epsfig{file=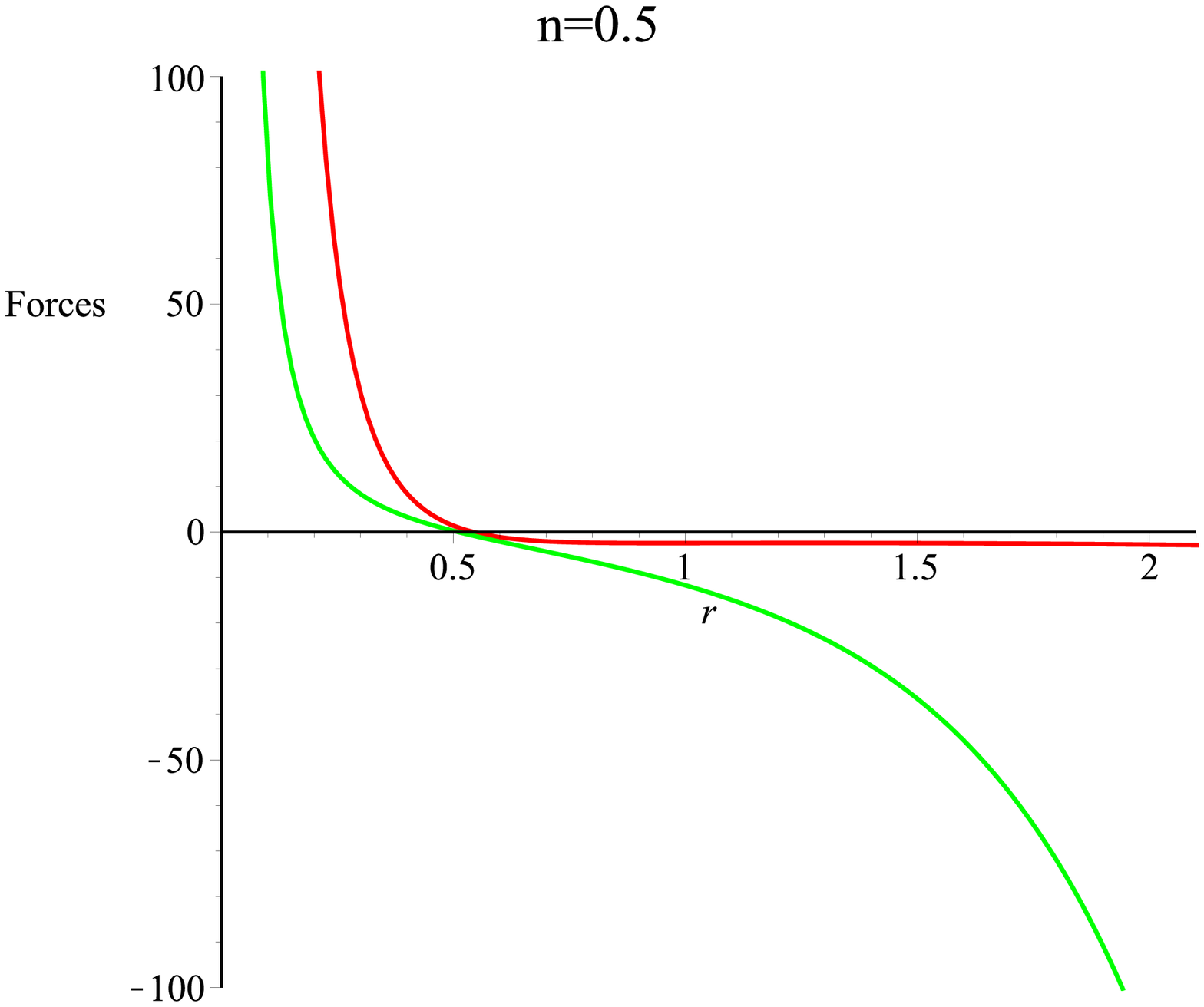,
width=0.4\linewidth}\epsfig{file=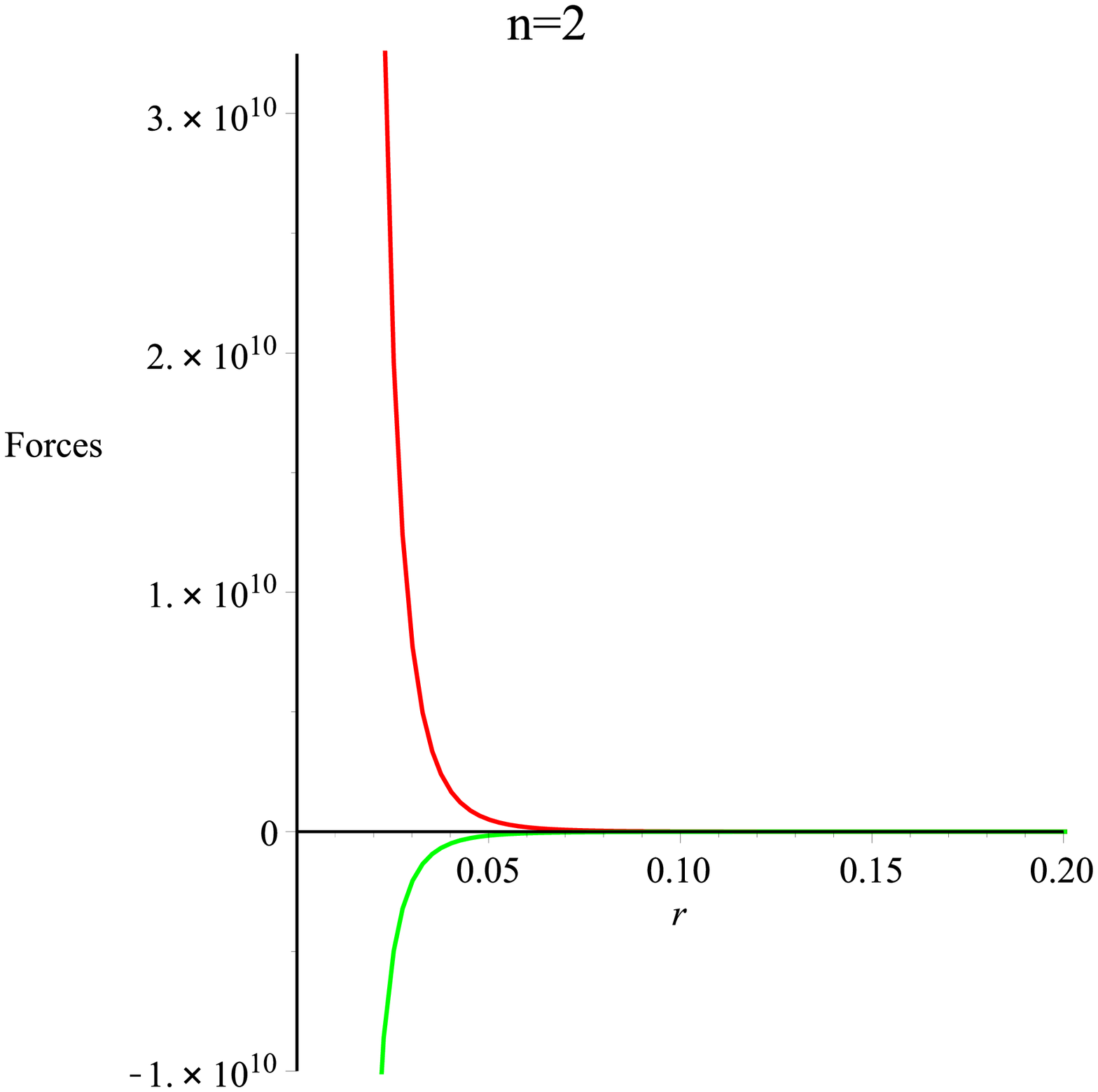,
width=0.4\linewidth}\\\epsfig{file=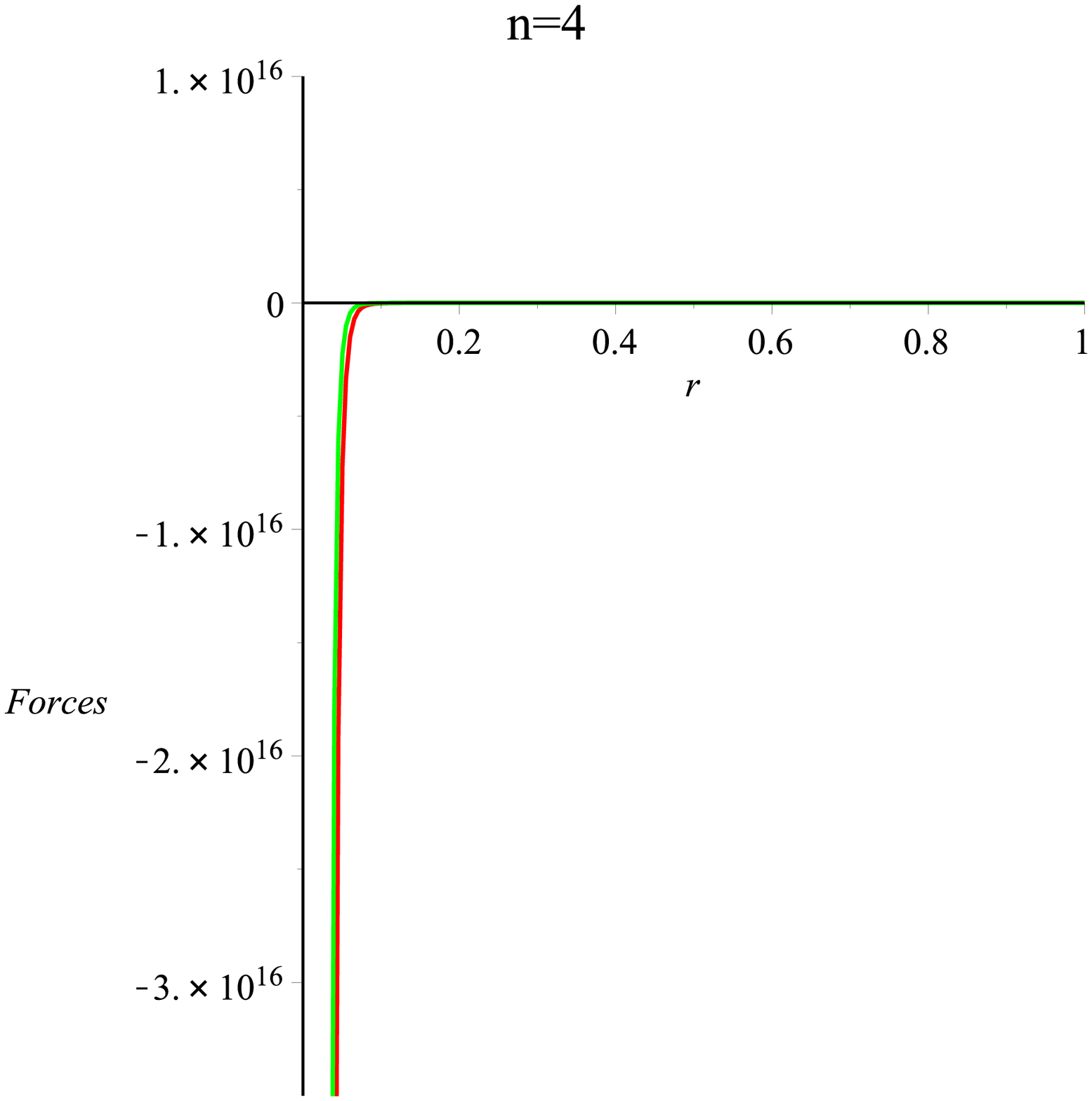,
width=0.4\linewidth}\caption{Plots of $\mathcal{F}_g$ (green) and
$\mathcal{F}_h$ (red) versus $r$ for $a(r)=k$, $d_{_2}=-2.2$,
$d_{_3}=1.001$, $d_{_4}=0.05$, $f_0=1$ and $\mathcal{M}_{eff}=2$.}
\end{figure}

In Figures \textbf{16}-\textbf{18}, we analyze the stability of WH
solutions constructed with the help of a new $f(R)$ model as well as
power-law and exponential forms of generic function $f(R)$. In
Figure \textbf{16}, the left plot represents the stability of WH
solution (\ref{35}) relative to constant red-shift function and
$f(R)$ model (\ref{A1}). The effect of gravitational and hydrostatic
forces appear to be the same but in opposite directions canceling
each other effect. Thus, the considered WH is found to be stable due
to null effect of these forces. For variable red-shift function, the
equilibrium state of WH solution (\ref{37}) is analyzed in the right
plot of Figure \textbf{16}. Initially, the WH geometry seems to be
unstable but gradually it attains an equilibrium state due to equal
but opposite effect of hydrostatic and gravitational forces. Figure
\textbf{17} determines the existence of stable WH for $n=0.5$, $n=2$
and $n=4$ with constant red-shift function. For $n=0.5$ and $n=0.4$,
the system remains unstable as $\mathcal{F}_g+\mathcal{F}_h\neq0$
whereas the constructed WH attains a stable state for $n=2$. In
Figure \textbf{18}, the WH solutions gradually attain equilibrium
state corresponding to both forms of red-shift function.
\begin{figure}\centering\epsfig{file=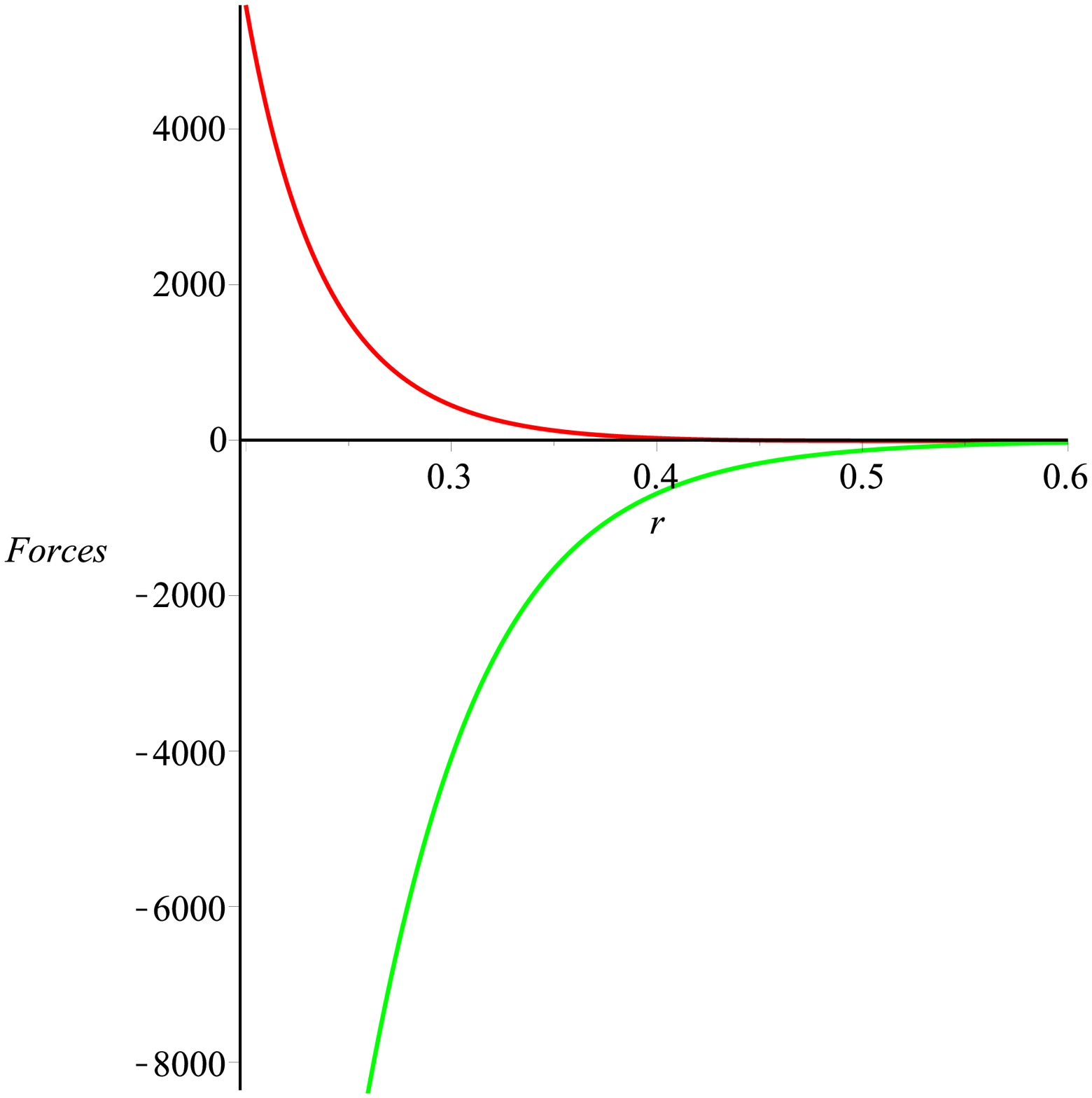,
width=0.45\linewidth}\epsfig{file=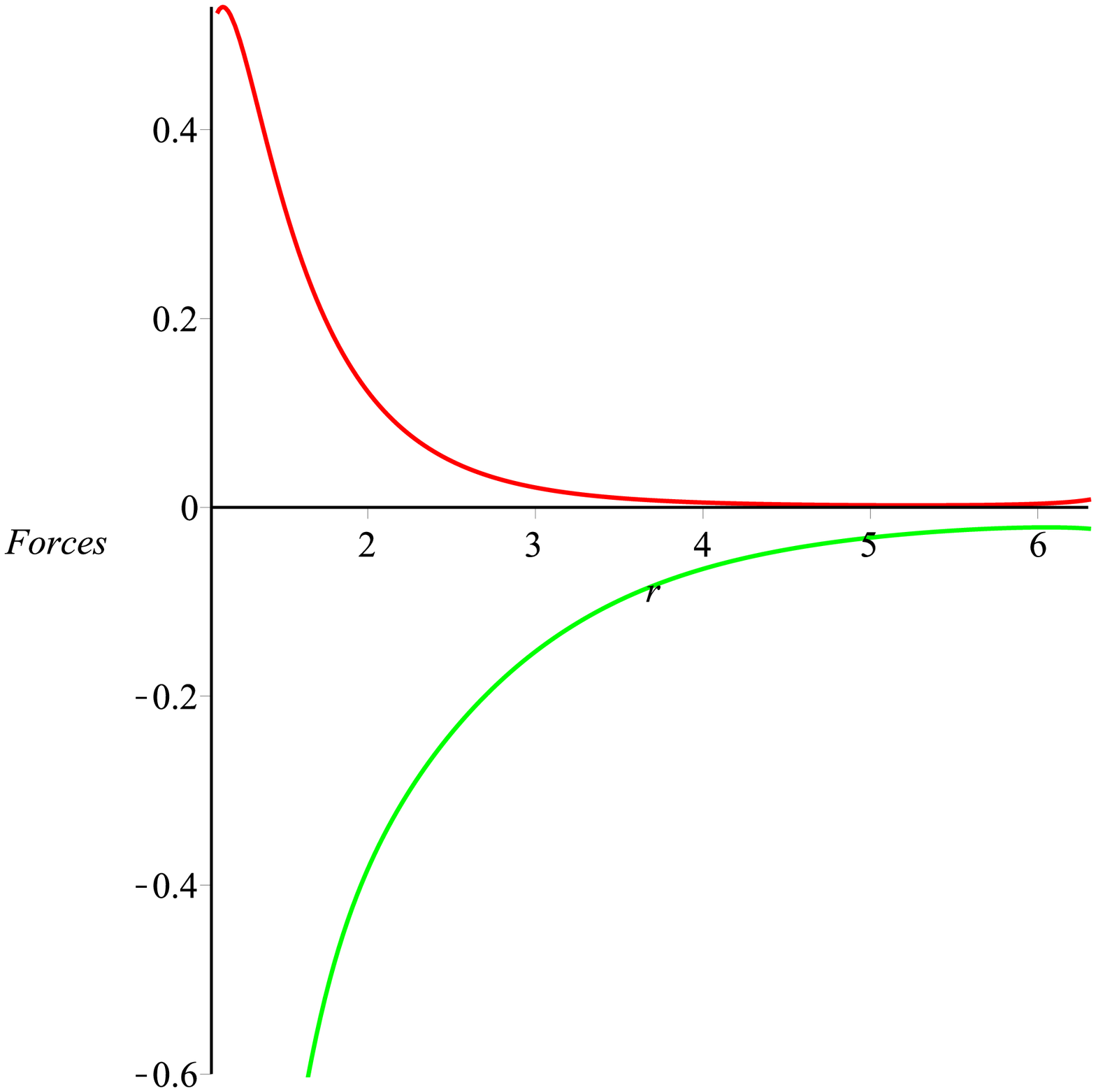,
width=0.45\linewidth}\caption{Plots of $\mathcal{F}_g$ (green) and
$\mathcal{F}_h$ (red) versus $r$ for $a(r)=k$ (left), $k=0.005$,
$\mathcal{M}_{eff}=-2$ and $a(r)=-k/r$ (right), $\chi_{_4}=-0.2$,
$R_{_0}=-0.95$, $k=2$ and $\mathcal{M}_{eff}=2$.}
\end{figure}

\section{Final Remarks}

In general relativity, the physical existence of a static
traversable WH demands the violation of NEC by the energy-momentum
tensor. This violation confirms the presence of exotic matter which
would be minimized to have a physically viable WH. In case of $f(R)$
gravity, the energy-momentum tensor threading WH satisfies NEC and
WEC whereas the existence of exotic matter is assured by the
effective energy-momentum tensor which violates NEC. In this paper,
we have discussed the presence of static traversable WH via Noether
symmetry approach in $f(R)$ gravity. For this purpose, we have
considered perfect fluid distribution and studied possible existence
of realistic WH solutions for generic as well as $f(R)$ power-law
model. We have solved over-determined system by invariance condition
and found symmetry generator, associated conserved quantity, exact
solution of $f(R)$ and $b(r)$ for static spherically symmetric
metric. For these solutions, we have studied WH geometry and also
investigated stable state of WH solutions via modified TOV equation
for the red-shift function when $a(r)=k,~-k/r$.

In case of constant red-shift function, we have obtained viable
$f(R)$ model and the shape function satisfies all the properties,
i.e., $h(r)>0$, WH geometry is found to be asymptotic flat and
$\frac{dh(r)}{dr}<1$ at $r=r_0$. The violation of NEC (using
effective energy-momentum tensor) assures the presence of repulsive
nature of gravity while existence of ordinary matter is supported by
verification of NEC and WEC relative to perfect fluid. When
$a'\neq0$, the $f(R)$ model preserves stability conditions for
$0<\omega<-0.08$ and the shape function has preserved all conditions
of traversable WH while $\rho_{eff}+p_{eff}<0$, $\rho_m+p_m\geq0$
and $\rho_m\geq0$ minimizing the presence of exotic matter due to
the presence of repulsive gravity. These energy bounds confirm the
presence of a realistic WH solution threaded by $T^{(m)}_{\mu\nu}$.
Consequently, we have found a physically viable WH solution for
$a'\neq0$. For both forms of red-shift function, the constructed WH
solutions attain an equilibrium state as
$\mathcal{F}_g+\mathcal{F}_h=0$.

We have also formulated symmetry generator, corresponding first
integral and WH solutions for $f(R)$ power-law model. When
$a'(r)=0$, we have established graphical analysis of traversable WH
conditions for $n=1/2,~n=2$ and $n=4$. In this case, the shape
function is found to preserve all conditions and
$\rho_{eff}+p_{eff}<0$ assures the violation of NEC identifying the
existence of exotic matter at throat. The consistent behavior of
$\rho_m\geq0$ and $\rho_m+p_m\geq$ indicate that the constructed
traversable WH is supported by ordinary matter. The stability
analysis of these realistic traversable WHs identifies that the WH
geometry would be stable only for $n=2$. For $a'\neq0$, we have
found a complicated form of the shape function. For exponential
$f(R)$ model, the WH geometry is discussed near the throat. The
shape of WH is found to be asymptotically flat for both constant as
well as variable forms of the red-shift function. The violation of
effective NEC and verification of NEC as well as WEC of ordinary
matter assure the presence of realistic traversable WH solutions.
The total effect of gravitational and hydrostatic forces identifies
equilibrium state of WHs in both cases.

The WH solutions are found in $f(R)$ gravity which is equivalent to
Brans-Dicke theory under a particular conformal transformation.
Coule \cite{aop3} established static unrealistic WH solutions in
Einstein frame of $f(R)$ theory. Nandi et al. \cite{aop6} examined
the possibility of static WH solutions in the background of both
Jordan and Einstein frames of Brans-Dicke theory. They found that
the non-traversable WH exists in the former frame whereas in the
latter frame, WH solutions do not exist at all unless energy
conditions are violated by hand. Furey and de Benedictis \cite{aop5}
discussed geometry of the WH solutions near the throat while
Bronnikov and Starobinsky \cite{aop4} claimed that the existence of
throat can be preserved under a conformal transformation. In
general, the back transformation from Jordan to Einstein frames does
not assure to get physical solutions. It has been even widely
demonstrated that passing from one frame to the other can completely
change the physical meaning as well as the stability of the
solutions \cite{aop36a}. Bahamonde et al. \cite{aop2a} observed the
presence of big-rip (type I) singularity in the Einstein frame of
$f(R)$ gravity while along back mapping, the universe evolution is
found to be singularity free.

In this paper, we have explored the existence of realistic and
stable traversable WH solutions in the Jordan frame representation
of $f(R)$ theory. It is worth mentioning here that the WH geometry
is discussed at the throat in case of standard power-law and
constructed $f(R)$ models whereas in case of exponential model, we
have analyzed the WH geometry near the throat. The presence of
repulsive gravity due to higher order curvature terms leads to
traversable WHs while the existence of ordinary matter confirms the
realistic nature of these traversable WH solutions in each case. For
$f(R)$ power-law model, the WH solutions are stable only for $n=2$
while stability is preserved for both exponential as well as
constructed $f(R)$ models. It would be interesting to analyze the
presence of these configurations in the Einstein frame where
contribution of scalar field may enhance the traversable nature as
it introduces anti-gravitational effects. On the other hand, the
back mapping of these frames may or may not ensure the presence of
stable as well as realistic traversable wormholes.

\vspace{0.25cm}

{\bf Acknowledgment}

\vspace{0.25cm}

This work has been supported by the \emph{Pakistan Academy of
Sciences Project}.

\end{document}